\documentclass[]{aa}
\usepackage{graphicx}
\usepackage{txfonts}
\newcommand{\epsfig}[1]{\resizebox{\hsize}{!}{\includegraphics{#1}}}
\newcommand{\figcap}[1]{\caption[Another figure]{#1}}
\newcommand{\AJ}[3]{{#1}, AJ, \vol{{#2}}, {#3}.}
\newcommand{\ApJ}[3]{{#1}, ApJ, \vol{{#2}}, {#3}.}

\newcommand{\AandA}[3]{{#1}, A\&A, \vol{{#2}}, {#3}.}

\newcommand{\MNRAS}[3]{{#1}, MNRAS\rm, \vol{{#2}}, {#3}.}

\newcommand{\Acta}[3]{{#1}, Acta Astr.\rm, \vol{{#2}}, {#3}.}

\newcommand{\PASP}[3]{{#1}, PASP\rm, \vol{{#2}}, {#3}.}

\newcommand{\vol}[1]{{\mbox{#1}}}

\newcommand{\FeH}{\mbox{[Fe/H]}\,}

\newcommand{\kms}{\mbox{$\mbox{km\,s}^{-1}$}\,}

\begin{document}
\title{BVRIJK light curves and radial velocity curves\\
         for selected Magellanic Cloud Cepheids
       \thanks{Based on data
          acquired at the Las Campanas Observatory, Chile,
          the Cerro Tololo Inter American Observatory, Chile,
          and the European Southern Observatory, Chile}
       $\!^{,}\!$
       \thanks{
               {Tables 10-27 are only available in
                    electronic form at the CDS via anonymous ftp
                    to cdsarc.u-strasbg.fr or via
                    http://cdsweb.u-strasbg.fr/A+A.htx
               }
         }
	}

\author{Jesper~Storm\inst{1}
\and 
	Bruce~W.~Carney\inst{2}
\and
	Wolfgang~P.~Gieren\inst{3}
\and
	Pascal~Fouqu\'e\inst{4,5}
\and
	Wendy~L.~Freedman\inst{6}
\and
	Barry~F.~Madore\inst{7}
\and
	Michael~J.~Habgood\inst{2}
}
\institute{Astrophysikalisches Institut Potsdam,
        An der Sternwarte 16, D-14482 Potsdam, Germany;
	e-mail: jstorm@aip.de
\and
        Univ. of North Carolina at Chapel Hill,
        Dept. of Physics and Astronomy,
        Phillips Hall, Chapel Hill,
        NC-27599-3255, USA;
	e-mail: bruce@physics.unc.edu
\and
	Universidad de Concepci\'on, Departamento de F\'{\i}sica,
	Casilla 160-C, Concepci\'on, Chile;
	e-mail: wgieren@coma.cfm.udec.cl
\and
	Observatoire de Paris, Section de Meudon DESPA F-92195 Meudon Cedex,
	France
\and
	European Southern Observatory, Casilla 19001, Santiago 19, Chile;
	e-mail: pfouque@eso.org
\and
	The Observatories,
	Carnegie Institution of Washington,
	813 Santa Barbara Street,
	Pasadena, California 91101, USA;\\
	e-mail: wendy@ociw.edu
\and
	California Institute of Technology,
	IPAC, MC 100-22,
	Pasadena, California 91125, USA;
	e-mail: barry@ipac.caltech.edu
	}

\offprints{J. Storm (AIP), e-mail: jstorm@aip.de}
\date{Received 5 August 2003 / Accepted 14 November 2003}

\abstract{
We present high precision and well sampled $BVRIJK$ light curves and radial
velocity curves for a sample of five Cepheids in the SMC.
In addition we present radial velocity curves for three Cepheids in the LMC.
The low metallicity ($\FeH \approx -0.7$) SMC stars have been
selected for use in a Baade-Wesselink type analysis to constrain the
metallicity effect on the Cepheid Period-Luminosity
relation. The stars have periods of around 15 days so they are similar
to the Cepheids observed by the Extragalactic Distance Scale Key Project
on the Hubble Space Telescope.
We show that the stars are representative of the SMC Cepheid population at that
period and thus will provide a good sample for the proposed analysis.
The actual Baade-Wesselink analysis are presented in a companion paper.

}

\maketitle

\section{Introduction}

  Recently Freedman et al. (\cite{Freedman01}) have concluded the Hubble Space
Telescope key project on the Extragalactic Distance Scale. They applied
the Cepheid Period-Luminosity relation to Cepheids in distant galaxies
to calibrate secondary standard candles which in turn can be used to
measure the free Hubble flow in the Universe to an accuracy of better
than 10\%. This result depends on the distance to the Large Magellanic
Cloud, which has been assumed to be $(m-M)_0 = 18.5\pm 0.1$, and
on the assumption that the
PL relation is only weakly dependent on metallicity. Many studies in the
last decade have tried to resolve both of these issues, but widely
differing results keep turning up.

  In this paper we present data which will be used for a study of the
metallicity effect on the PL relation. The idea is to determine the
luminosities of individual, low metallicity ($\FeH \approx -0.7$),
SMC Cepheids and compare these luminosities to those of
solar metallicity Galactic Cepheids of similar period.
The low metallicity of the SMC
Cepheids combined with the solar metallicity Galactic stars provide a
reasonably large range in metallicity to place significant constraints
on the effect metallicity has on the PL relation.

  The recent calibration of the infrared surface brightness
method (Fouqu\'e \& Gieren \cite{Fouque97}) provides
a powerful tool for determining luminosities to individual Cepheids.
However, to exploit the full potential of the method accurate optical
($V$-band) and near-IR ($K$-band) light curves, as well as well-sampled
and accurate radial velocity curves are needed. In the present paper we
present all the observational data which is needed for the analysis.
Preliminary analyses and discussions have already been presented in
Storm et al. (\cite{Storm98}, \cite{Storm99} and \cite{Storm00}), and
in a companion paper (Storm et al. \cite{Storm03}) we present our final
surface brightness analysis of the stars and determine the constraints on
the metallicity effect on the Cepheid PL relation, and the consequences
for the value of the Hubble constant.

  Five Cepheids in the south-western part of the SMC about 1 degree from
the main body of the galaxy were selected for further study. With this
location the stars are representative of the SMC population but not
affected by the variable extinction and severe crowding which is found
closer to the main body of the SMC.
The coordinates for the stars are listed in Table \ref{tab.coords}. 
Finding charts can be found in Hodge \& Wright (\cite{Hodge77}). The
stars were selected to have periods around 15 days, long enough to avoid
overtone pulsators, and to guarantee similarity to Cepheids used for
extragalactic
distance determination, but still short enough to allow a complete phase
coverage with a reasonable number of nights at the telescope. Using a sample
of almost equal period stars also allows us to probe the
intrinsic width of the instability strip at this period.

  In addition to the SMC sample we also observed a few LMC Cepheids of
different periods, and for completeness we present the data acquired for
these stars here as well.

\begin{table}
\caption{\label{tab.coords} The coordinates of the Cepheids as returned by
skycat from the Digital Sky Survey (J2000.0).}
\begin{tabular}{l l c c}
\hline\hline
Identifier & OGLE ID & RA & DEC \\
 & & hh:mm:ss & dd:mm:ss \\
\hline
\multicolumn{4}{c}{\em SMC} \\
  HV~822 & SC2-84516 & 00:41:55.6 & $-73$:32:25 \\
 HV~1328 &           & 00:32:54.7 & $-73$:49:20 \\
 HV~1333 &           & 00:36:03.5 & $-73$:55:58 \\
 HV~1335 & SC1-00001 & 00:36:55.4 & $-73$:56:30 \\
 HV~1345 & SC2-41552 & 00:40:38.5 & $-73$:13:14 \\
\multicolumn{4}{c}{\em LMC} \\
 HV~2694 & SC16-70661 & 05:35:32.0 & $-69$:43:22 \\
HV~12797 &           & 05:22:24.0 & $-71$:59:19 \\
HV~12198 &           & 05:13:26.7 & $-65$:27:05 \\
\hline
\end{tabular}
\end{table}


\section{The Optical Photometric Data}

\subsection{The Observations}

  The Cepheids were observed with a number of CCD cameras
at the Las Campanas, Cerro Tololo and La Silla observatories in Chile
as summarized in Table \ref{tab.calcoeff}.
The stars were observed in the $BVR$ and $I$ bands and the exposure times
ranged from 30 secs to a few minutes depending on the instrument
sensitivity, the seeing and the transparency of the sky.
The seeing was typically in the range 0.9 to 1.4~arcsec and all frames
with seeing worse than about 2~arcsec were discarded as crowding effects
started to become significant. In all cases the point spread function (PSF) was
well sampled by the detector.

\subsection{Data reduction}

\subsubsection{Pre-processing}

  The CCD frames were bias subtracted using a constant
value for the bias level determined from bias exposures. The flat
fielding was done using sky flatfields for the large scale
structure and dome flatfields for the pixel to pixel variations in case
of the Las Campanas data. For the CTIO data, the high read-out speed of
the detector (about 15 seconds for a full frame) allowed the
exclusive use of sky flatfields for the gain calibration.

The 1989 observing run at Las Campanas was somewhat compromised by
significant gain variations of the UV-flooded Texas Instrument CCD
during the individual nights.  After various experiments we have been
reasonably successful in scaling and interpolating the flatfields for
each science frame to compensate the effect, but these data are of
lower quality than those obtained with the other instruments.

\subsubsection{Relative photometry}

  The relative photometry was determined using the DoPHOT\_2.0 package
(Schecter et al. \cite{Schecter93}).
The intermediate photometric system was defined by the CTIO
observations from the night Jan.2, 1994 as the CTIO system is close
to the standard Landolt (\cite{Landolt92}) system. Furthermore the seeing and
photometric quality was good on this night and seven standard field
observations were performed (see below).

  For each of the other observing runs transformation to the CTIO system
of the form
\begin{eqnarray}
    v_{\mbox{\scriptsize ctio}} & = & v_{\mbox{\scriptsize
inst}} + \alpha_{vbv}\times (b-v)_{\mbox{\scriptsize inst}} + \mbox{const} \label{eq.vbv}\\
(b-v)_{\mbox{\scriptsize ctio}} & = & \alpha_{bv}\times (b-v)_{\mbox{\scriptsize
inst}} + \mbox{const}\\
(v-r)_{\mbox{\scriptsize ctio}} & = & \alpha_{vr}\times (v-r)_{\mbox{\scriptsize
inst}} + \mbox{const}\\
(v-i)_{\mbox{\scriptsize ctio}} & = & \alpha_{vi}\times (v-i)_{\mbox{\scriptsize
inst}} + \mbox{const}\label{eq.vivi}
\end{eqnarray}
\noindent
were determined. The
transformations were first determined for each set of instrument
magnitudes $bvri$ based on the brightest (lowest estimated error) stars
in common with the CTIO data. Typically more than 100 stars in each field
were used.  The coefficients were then averaged over each observing run
giving the values listed in Table \ref{tab.calcoeff}.  The typical RMS for
the color terms is of the order $0.02$~mag.  These averaged coefficients
were then used for the next iteration where only the zero points for
the individual frames were fitted. The final transformation of the $bvri$
sets to the CTIO system was then performed using these zero points and
the average transformation coefficients.

\begin{table*}
\caption{\label{tab.calcoeff} Coefficients for the transformation of the
various instrument systems to the CTIO instrument system.}
\begin{tabular}{r l l r r r r}
\hline\hline
Observing period & Observatory & Instrument (Detector) & $\alpha_{vbv}$ & $\alpha_{bv}$ &
$\alpha_{vr}$ & $\alpha_{vi}$ \\
\hline
Nov. 1988 & LCO & 1m (TI\#1) & $-0.089$ & 1.168 & 0.890 & 0.898 \\
Jul. 1989 & CTIO & 0.9m (RCA\#5)& $-0.077$ & 1.177 & 0.963 & 0.963 \\
Jul. 1989 & CTIO & 0.9m (TI\#3) & $-0.078$ & 1.112 & 0.933 & 0.945 \\
Aug. 1989 & CTIO & 4m (TI\#1) & $-0.071$ & 1.268 & 0.943 & 0.937 \\
Nov. 1989 & LCO & 1m (TI\#2) & $-0.089$ & 1.168 & 0.890 & 0.898 \\
Dec. 1993 & CTIO & 0.9m (Tek\#2) & $0.000$ & 1.000 & 1.000 & 1.000 \\
Jan. 1996 & ESO & DFOSC (C1W11/4) & $0.032$ & 1.227 & 0.928 & 1.022 \\
Jul. 1996 & ESO & DFOSC (C1W11/4) & $0.032$ & 1.227 & 0.928 & 1.022 \\
Jul. 1996 & ESO & EFOSC-2 (CCD\#40) & $0.066$ & 1.076 & 1.023 & 1.025 \\
Sep. 1996 & ESO & DFOSC (C1W11/4) & $0.017$ & 1.097 & 0.939 & 1.021 \\
Oct. 1996 & CTIO & 0.9m (Tek\#3)  & $-0.009$ & 0.932 & 0.989 & 0.982 \\
\hline
\end{tabular}
\end{table*}

\subsubsection{Transformation to the standard system}

  On the night of Jan. 2, 1994 we observed three Landolt (\cite{Landolt92}) fields.
Each field was observed two or three times at different airmass
giving a total of 31 $BVRI$ measurements of standard stars
covering a wide range in color ($-0.3 < (B-V) < 1.9$) and airmass
($1.1< X < 1.8$). Synthetic aperture photometry was performed using
the IRAF\footnote{IRAF is the Image Reduction and Analysis
Facility, made available to the astronomical community by the National
Optical Astronomy Observatories, which are operated by AURA, Inc., under
cooperative agreement with the National Science Foundation.}
{\tt noao.digiphot.apphot} package and employing an aperture radius of 18
pixels corresponding to a diaphragm diameter of 14 arc-seconds, which also
happens to be very similar to the one employed by Landolt (\cite{Landolt92}).

  The IRAF {\tt photcal} package was used to derive the transformations
in the form

\begin{equation}
M = m + \alpha_m \times \mbox{color} + k_m \times X + c_m
\end{equation} 

\noindent
where $M$ refers to standard magnitudes and $m$
to the CTIO instrumental magnitudes, color refers to the instrumental
color term ($(b-v)$
or $(v-r)$ or $(v-i)$) as listed in
Table \ref{tab.stdcoeff}, $k_m$ is the extinction coefficient for
passband $m$, $X$ is the airmass and $c_m$ is the zero point for
passband $m$.

  The values of the coefficients with the estimated errors are listed in
Table \ref{tab.stdcoeff}. Note that the CTIO instrument system closely
resembles the standard system.

\begin{table}
\caption{\label{tab.stdcoeff} Coefficients for the transformation of the
CTIO instrument system to the Landolt (\cite{Landolt92}) system.}
\begin{tabular}{c r r r r}
\hline\hline
  & \multicolumn{1}{c}{$B$} & \multicolumn{1}{c}{$V$} &
  \multicolumn{1}{c}{$R$} & \multicolumn{1}{c}{$I$} \\
\hline
color & $(b-v)$ & $(b-v)$ & $(v-r)$ & $(v-i)$ \\
$\alpha_m$ & 0.007 & 0.019 & 0.008 & 0.006 \\
$\sigma_\alpha$ & 0.006 & 0.003 & 0.005 & 0.003 \\
$k_m$ & $-0.249$ & $-0.147$ & $ -0.118$ & $-0.059$ \\
$\sigma_k$ & 0.013 & 0.007 & 0.007 & 0.009 \\
$c_m$ & $-4.913$ & $-4.204$ & $-4.183$ & $-5.013$ \\
$\sigma_c$ & 0.018 & 0.010 & 0.010 & 0.012 \\
RMS & 0.018 & 0.012 & 0.010 & 0.012 \\
\hline
\end{tabular}
\end{table}

To tie in the instrumental magnitudes from the DoPHOT PSF fitting
photometry, $m_{\mbox{\scriptsize PSF}}$, with the aperture photometry
of the standard star observations we determined the aperture corrections
for the PSF photometry for each individual field.  \object{HV~822} was
the only field which was not observed during this night, and it was tied
in with the \object{HV~1345} field on the following night using a similar
procedure.


In the case of \object{HV~1335} it was necessary to add an offset of
$\Delta V = -0.1$~mag to the magnitudes from the Nov. 1989 run to
bring the $V$ light curve into agreement with that obtained from the
other instruments.  This is the only deviation from the transformations
given above, and it is justified as some of the frames from this run are likely
to suffer from residual systematic errors in the flatfielding. The fact that the
other stars agree reasonably well with the measurements from the
other runs indicate that the flatfielding for the Nov. 1989 run overall
has been quite successful.

\subsection{Ephemerides}

The periods for the stars were re-derived from the present data by
minimizing the dispersion between the points in the V-band light curve.
The epoch of maximum $V$-light was finally determined by shifting the light
curve by eye to obtain a good fit. For \object{HV~1335} we have adopted
the values found by Udalski et al. (\cite{Udalski99}).  The adopted
ephemerides are listed in Table \ref{tab.ephem}.

\begin{table}
\caption{\label{tab.ephem} Ephemerides adopted for the stars.}
\begin{tabular}{l c c l}
\hline\hline
Identifier & Period & $\log_{10}(P)$ & Epoch \\
 & days & & HJD \\
\hline
\object{HV~822} & 16.7421 & 1.223810 & 2447485.9 \\
\object{HV~1328} & 15.8360 & 1.199645 & 2447486.7 \\
\object{HV~1333} & 16.2935 & 1.212014 & 2447491.3 \\
\object{HV~1335} & 14.3816 & 1.157807 & 2450610.6 \\
\object{HV~1345} & 13.4784 & 1.129638 & 2447496.0 \\
\object{HV~2694} & 6.9363 & 0.841128 & 2449349.4 \\
\object{HV~12797} & 6.82   & 0.8338 & 2432000.0\\
\object{HV~12198} & 3.5228 & 0.550573 &2432011.6 \\
\hline
\end{tabular}
\end{table}

\subsection{The light curves}

\begin{figure*}[htp]
\epsfig{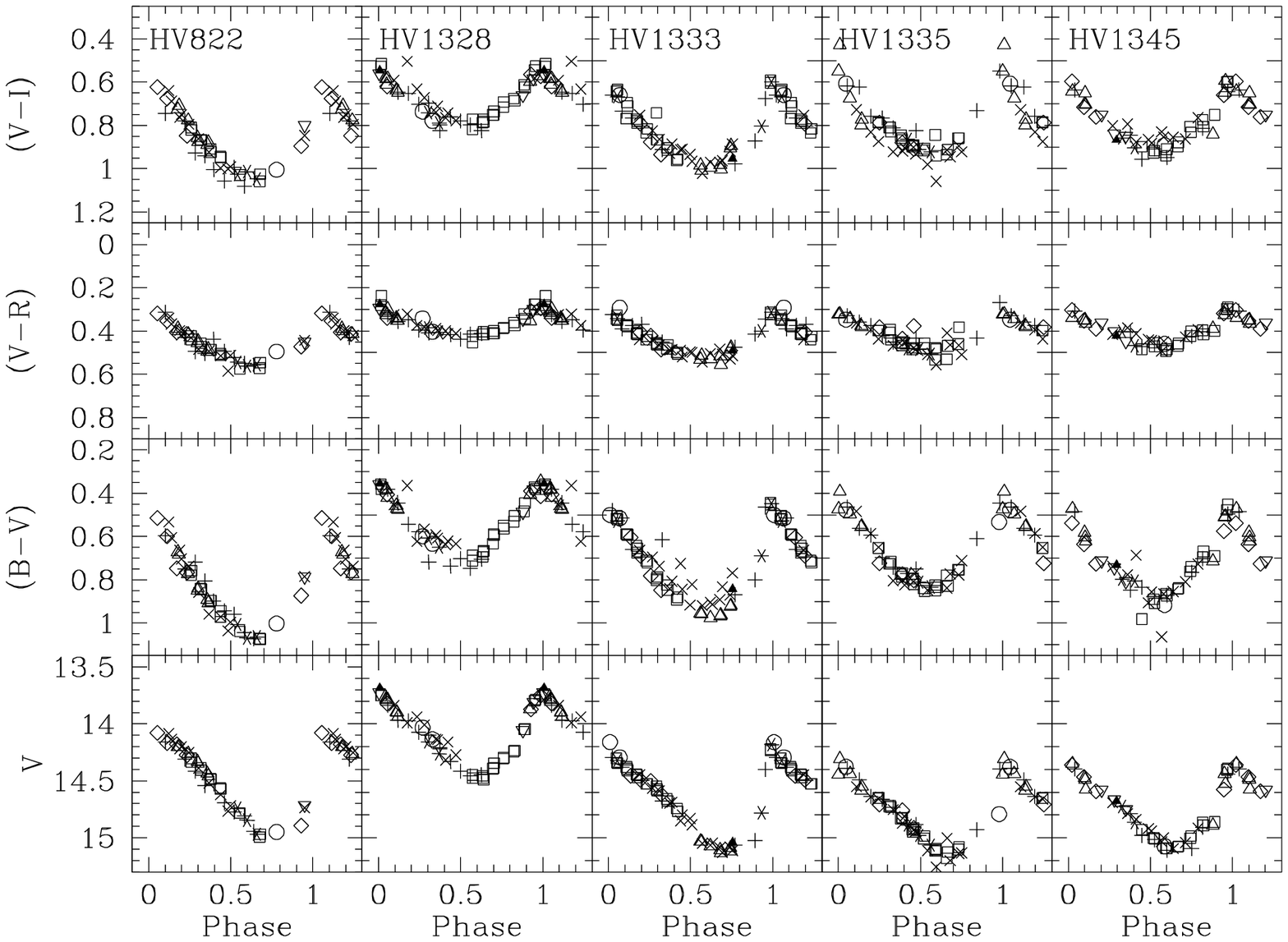}
\figcap{\label{fig.bvriall} Light and color curves for the SMC Cepheids. 
The data are marked in the following way: 
plusses: LCO~1988,
crosses: LCO~1989,
open diamonds: CTIO~1989,
open squares: CTIO~1993,
open inverse triangles: DFOSC Jan.~1996,
filled triangles: DFOSC Jul.~1996,
open circles: EFOSC-2 Jul.~1996,
open triangles: DFOSC Sep.~1996,
asterisks: CTIO 1996.}
\end{figure*}

  The photometry for the stars is given in Table
\ref{tab.HV822}-\ref{tab.HV1345} together with the photometric errors as
returned by DoPHOT. The light curves are plotted in
Fig.\ref{fig.bvriall} where the different observing runs
are indicated by a different symbol. The phases were determined on
the basis of the ephemerides from Table \ref{tab.ephem}.
  
\subsection{Comparison with other sources of photometry}
\label{sec.photcomp}

Caldwell \& Coulson (\cite{CC84}) presented photoelectric photometry for
\object{HV~1335}.  We find excellent agreement with their light curve when
we add an offset of $+0.07$~mag to their $V$ magnitudes. Such an offset
is expected as \object{HV~1335} has two faint ($\Delta V \approx 3.5$)
companions within a distance of 5~arcsec which will make the aperture
photometry (which includes the two companions) brighter by approximately
the observed amount.

Laney \& Stobie (\cite{LS94}) quote mean $V$ magnitudes for the
two stars \object{HV~1328} and \object{HV~1335} but again based on
photoelectric measurements so a precise agreement cannot be expected due
to contamination within either the object aperture or the sky aperture.

More recently Udalski et al. (\cite{Udalski99}) (The OGLE SMC Cepheid
database) have published CCD-based $BVI$ light curves for three of
the stars presented here, namely \object{HV~822}, \object{HV~1335} and
\object{HV~1345}.  For each of these stars they have about 10 $B$-band,
25 $V$-band, and 150 $I$-band observations with good phase coverage. The
agreement with the photometry presented here is good with the zero
points agreeing to better than 0.01~mag suggesting that our photometric
zero-point is accurate to about 0.01~mag. The only offsets which we have
deemed necessary are in the $I$-band where we have
offset their photometry for HV~822 by $+0.02$~mag and for HV~1335 by
$+0.03$~mag before we determined the mean value.

\begin{figure}[htp] \epsfig{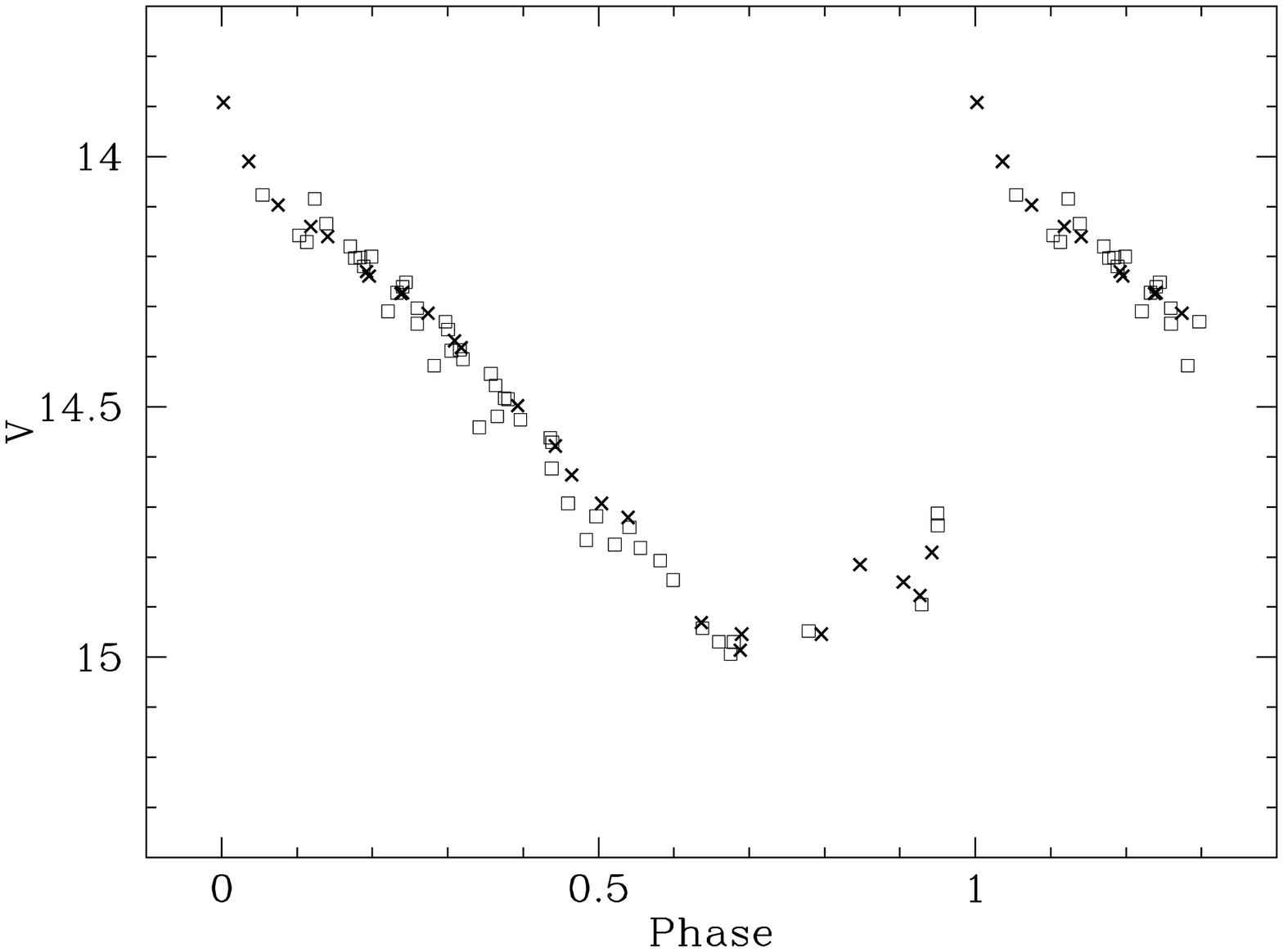}
\figcap{\label{fig.usOGLE822} The $V$-band light curve for \object{HV~822}
(open boxes) from our data with the measurements from 
Udalski et al. (\cite{Udalski99}) overplotted (crosses).}
\end{figure}

In Fig.\ref{fig.usOGLE822} the OGLE V-band light curve for \object{HV~822} has
been overplotted on our measurements. The agreement is excellent
even though the intensity means of the two data sets differ by 0.05~mag.
This difference is,
however, mainly due to a different phase coverage especially in the
phase range 0.8 to 1.0 where the magnitudes change rapidly and where
the light curve exhibits several inflection points.

\section{The near-IR data}

  $J$ and $K$-band data were acquired for the five SMC stars with a
variety of instruments as tabulated in Table \ref{tab.IRlog}. As the
period of the stars are so long, it only made sense to obtain one or at
most two phase points per night. With the limited number of stars in the
sample it was possible to obtain these data within a short time in
parallel to the main observing program (Cepheids in young LMC clusters).

\begin{table}
\caption{\label{tab.IRlog}List of observing runs which contributed
to the near-IR data set.}
\begin{tabular}{r l l }
\hline\hline
Date & Observatory & Instrument \\
\hline
Jan. 1998     & ESO, La Silla & 2.2m with IRAC-2\\
Oct. 1998     & Las Campanas & 1.0m with C40IRC \\
Dec. 1998     & Las Campanas & 2.5m with IRCAM \\
Jan. 1999     & Las Campanas & 1.0m with C40IRC \\
Jan. 1999     & Las Campanas & 2.5m with IRCAM \\
Jan. 2000     & CTIO & 1.5m with OSIRIS \\
\hline
\end{tabular}
\end{table}

\subsection{Data reduction}

  The observations were made in a five
position dither mode enabling a sky frame to be constructed by median
filtering these dithered frames. This local sky was subtracted from the
individual science frames which were then flatfielded by an appropriate
flatfield. The flatfield was either made from twilight exposures with an
appropriate dark frame subtracted (Las Campanas) or from dome flats 
where high- and low-light intensity frames with identical exposure time
were subtracted from each other (ESO and CTIO). 

A linearity correction was
applied to the raw frames where appropriate, but all science and calibration
frames were obtained with sufficiently low count levels, so the 
non-linearity does not affect the data ($< 1\%$ correction).

\subsection{Photometry}

  The photometry was done using the PSF fitting program DAOPHOT-II
(Stetson, \cite{Stetson87}) within IRAF. An instrumental photometric system was
defined for each field by a set of exposures obtained on a night of good
photometric quality. All the frames for the field from the other nights
and other instruments were transformed to this system in a manner
similar to the process described in a previous section for the optical data.
Color terms were determined between the instrumental system and the
other instruments on the basis of photometry of stars in the young LMC
cluster NGC~2136 with a wide range of $(J-K)$ color. In all cases no
significant color term between the instrument systems could be
established at the 2\% level.

\subsection{Calibration to the standard system}

During the nights Dec.27 and Dec.28 1998, we observed eight standard stars
several times for a total of 21 standard star observations. The stars
were all from the list of Persson et al. (\cite{Persson98}) and span the
color range from $(J-K)=0.24$ to $0.95$.  Each observation consisted of
four dithered exposures in each of the $J$ and $Ks$ filters. The airmass
terms were determined to be similar to the canonical values ($k_J=0.10$,
$k_K=0.08$) adopted by Persson et al. (\cite{Persson98}) so we adopted
these values. We also, as expected, did not detect a color term with
respect to the Persson system so we only had to determine the nightly
zero points. For the reddest measurements ($(J-K)>0.45$) we applied
the transformation determined by Persson et al. (\cite{Persson98}) to
transform the data to the CIT system, otherwise the two systems were
considered identical.

  On the same nights the science fields were also observed, and tertiary
reference stars in the fields were tied in with the standard system by
applying the offsets determined for the standard stars and correcting
for the appropriate airmass term. This was done using the IRAF package
{\tt photcal} on synthetic aperture photometry of isolated
stars in the fields and using growth curves to take out any variation
which might be present due to seeing variations.

  The observations from the other
nights were offset to the Las Campanas instrument system as described in
the previous section and these transformed data were finally transformed
to the CIT system using the zero points offsets determined here.
The resulting photometry is tabulated in Table
\ref{tab.HV822IR}-\ref{tab.HV1345IR}.

\begin{figure}[htp]
\epsfig{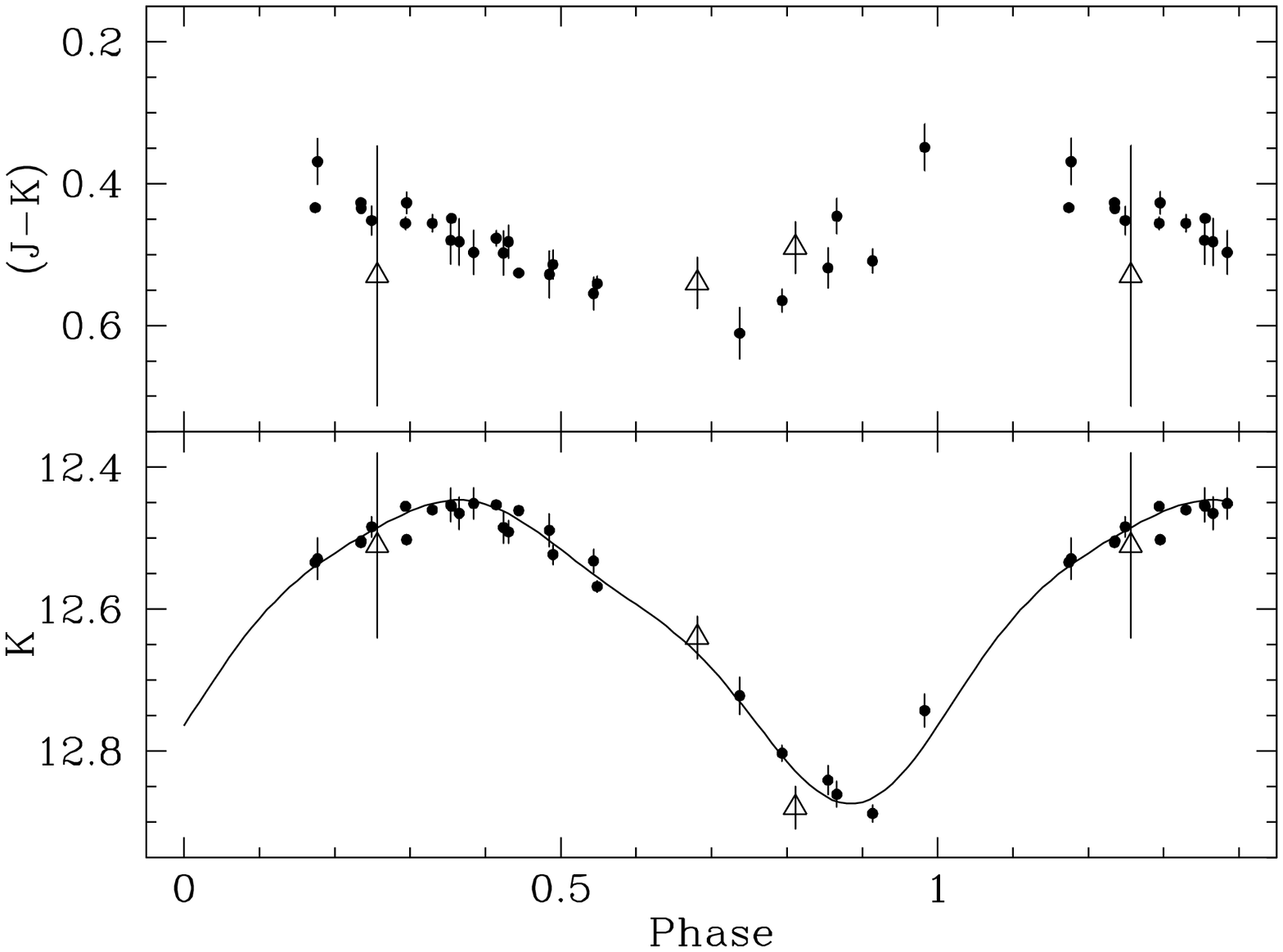}
\centering
\figcap{\label{fig.ir822} Near-IR light and color curves for \object{HV~822}. Open
triangles are from Welch et al. (\cite{Welch87}).}
\end{figure}

\begin{figure}[htp]
\epsfig{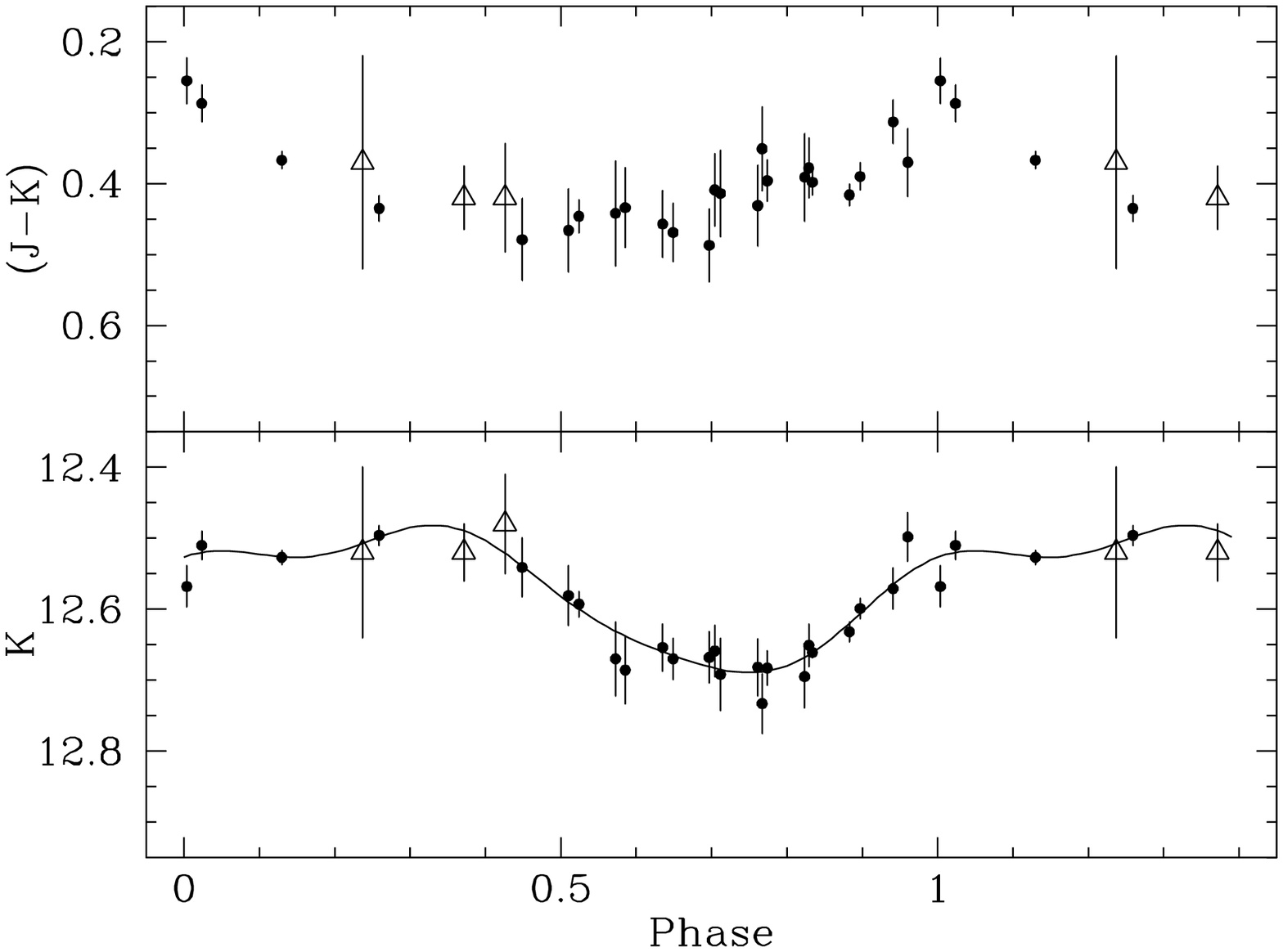}
\centering
\figcap{\label{fig.ir1328} Near-IR light and color curves for
\object{HV~1328}. Open
triangles are from Welch et al. (\cite{Welch87}).}
\end{figure}

\begin{figure}[htp]
\epsfig{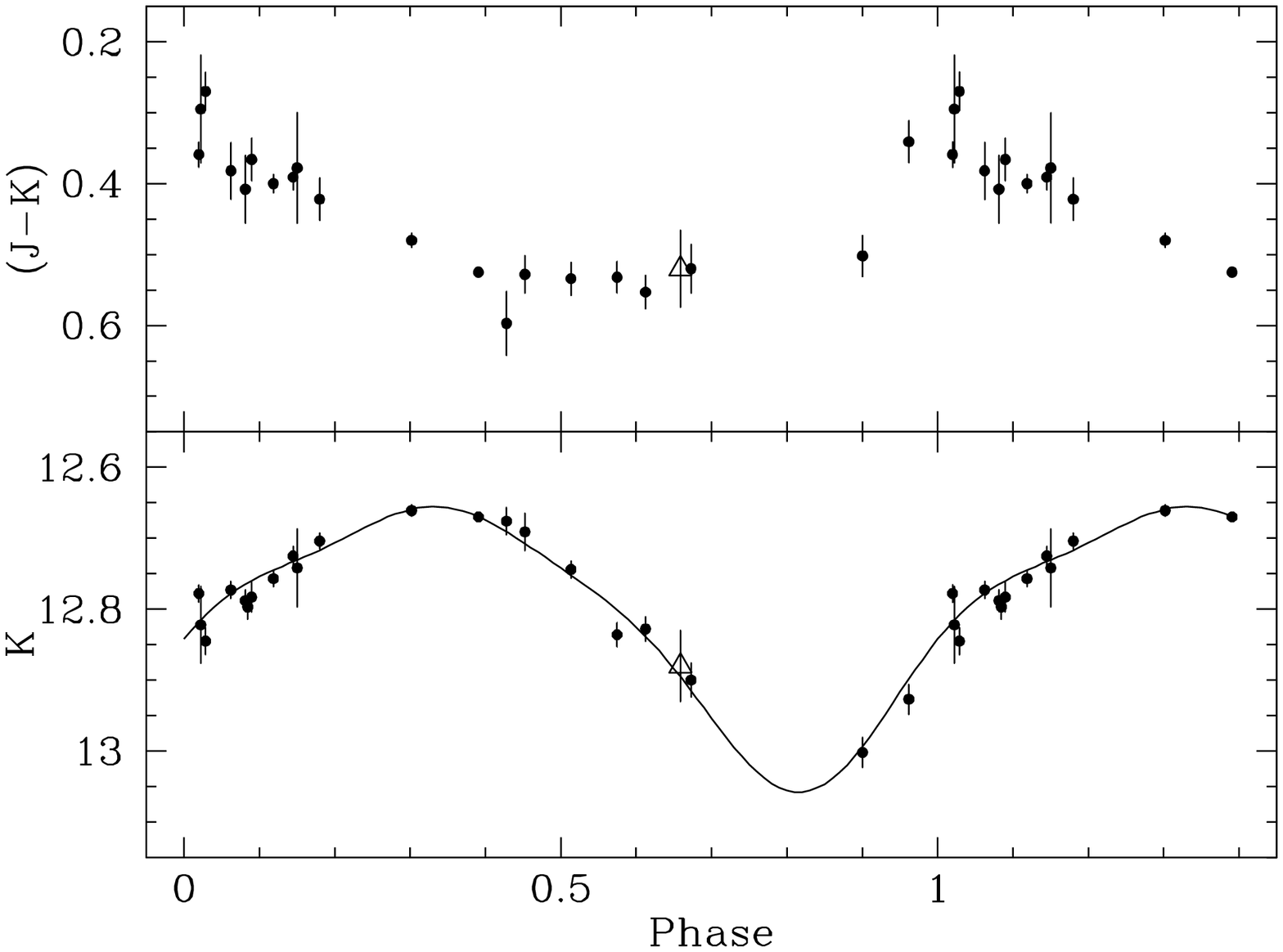}
\centering
\figcap{\label{fig.ir1333} Near-IR light and color curves for
\object{HV~1333}. Open
triangles are from Welch et al. (\cite{Welch87}).}
\end{figure}

\begin{figure}[htp]
\epsfig{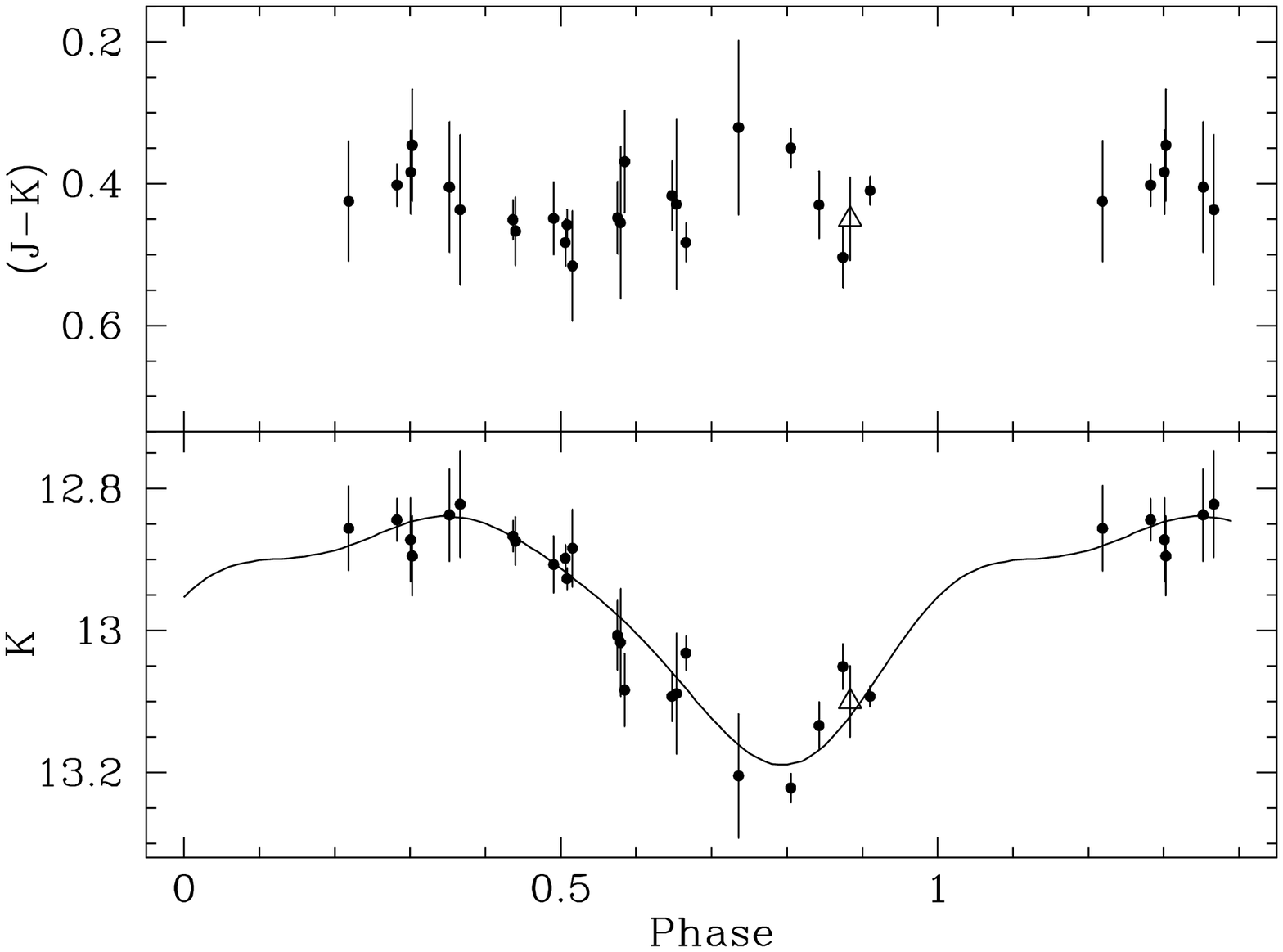}
\centering
\figcap{\label{fig.ir1335} Near-IR light and color curves for
\object{HV~1335}. Open
triangles are from Welch et al. (\cite{Welch87}).}
\end{figure}

\begin{figure}[htp]
\epsfig{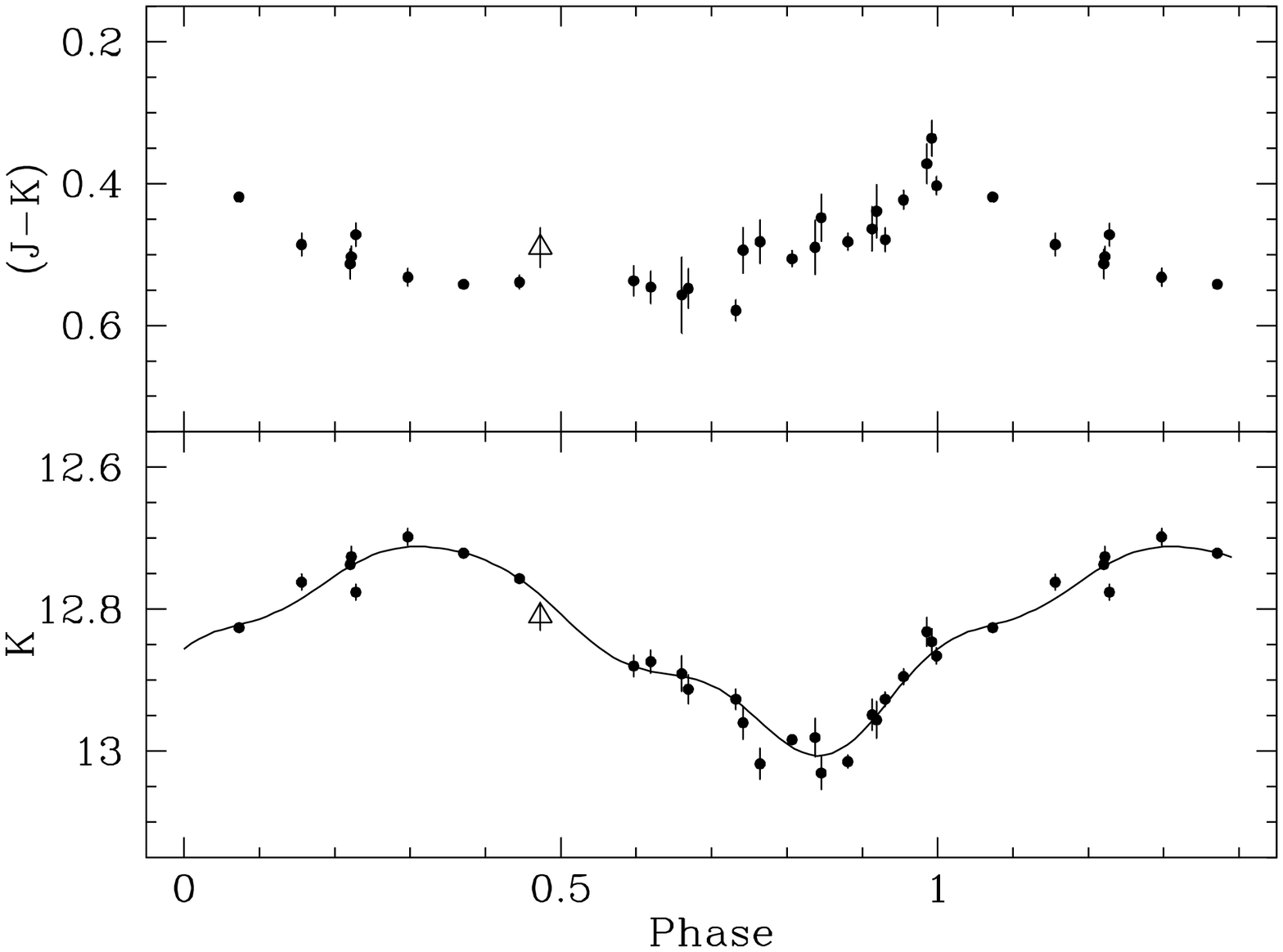}
\centering
\figcap{\label{fig.ir1345} Near-IR light and color curves for
\object{HV~1345}. Open
triangles are from Welch et al. (\cite{Welch87}).}
\end{figure}

\subsection{The $K$-band light curve and $(J-K)$ and $(V-K)$ color curves}

  The main purpose of the near-IR observations is to obtain a well
defined $(V-K)$ color curve for each Cepheid 
which is necessary for the application of
the near-IR surface brightness relation to these stars. At the same time
we can obtain good $J$ and $K$ light curves and thus a good estimate of
the mean $J$- and $K$-band magnitudes of the stars, for comparison with the 
corresponding Cepheid PL relations. The $(J-K)$ color curve can
potentially also be used for a Baade-Wesselink type analysis, but for
these stars it turns out that the amplitude is very small and thus the
photometric uncertainties become very significant. In short, the $(J-K)$
color is a poor (insensitive) temperature indicator for these stars.

  We have compared our $K$-band photometry with the measurements by
Welch et al. (\cite{Welch87}) and find excellent agreement. Their (few)
points have been
overplotted in the $K$-band light curves and will be used in the
determination of the $(V-K)$ color curve.

  As the near-IR data were obtained at different times from the $V$-band
observations
it is necessary to construct the $(V-K)$ color curve on the basis of the
$V$ and $K$-band light curves, i.e. in phase space. As the observations
are not obtained at exactly the same phases, it is necessary to interpolate
to obtain the color at a given phase. We decided to do this by fitting a
3rd or 4th order Fourier series to the $K$-band data and then use this
expansion to compute the
corresponding $K$ value for each $V$ phase point. The $K$-band light
curve has the lowest amplitude ($\pm0.15$~mag) and is significantly more
sinusoidal than the $V$-band light curve and thus lends itself well to a
low order Fourier fit. The low order fits follow the general trend of the
light curves rather well and the possible errors introduced from the fit
are small. Of course there is an enhanced uncertainty where the phase
coverage is poor. We have tried to use different (reasonable) orders for
the fits but the resulting Baade-Wesselink results were not affected
significantly. The Fourier fit to the data
points has been overplotted in Fig.\ref{fig.ir822}-\ref{fig.ir1345}. 

  To obtain the best possible phase coverage we have combined our
$V$-band data with those of Udalski et al. (\cite{Udalski99}). No offsets
have been applied as we could not determine any offset with a
significance of more than 0.01~mag.

\begin{figure*}[htp]
\epsfig{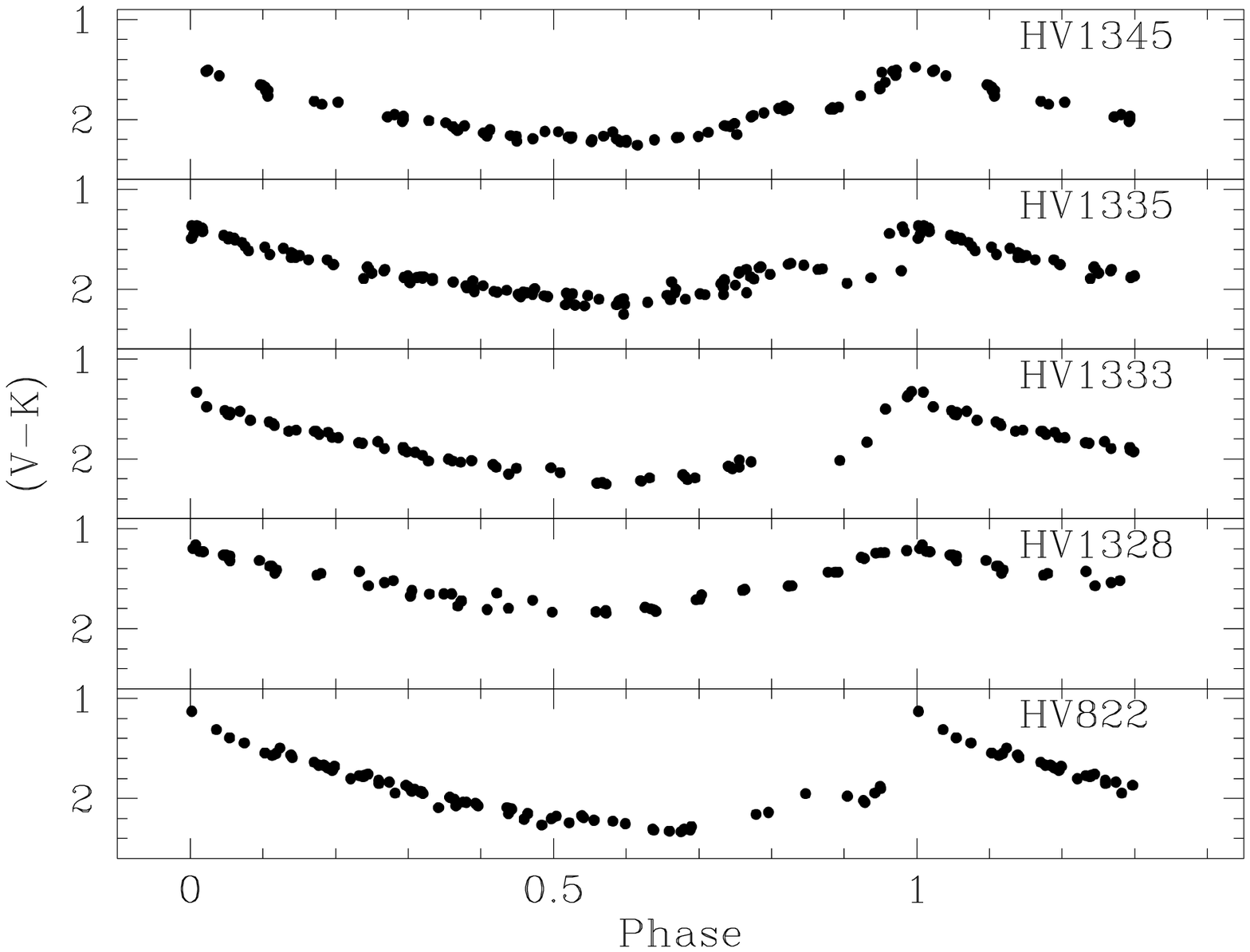}
\figcap{\label{fig.vkall} The $(V-K)$ color curves for the five prime
targets.}
\end{figure*}

  The resulting $(V-K)$ color curves are shown in Fig.\ref{fig.vkall}
and the photometric values are tabulated in Table
\ref{tab.HV822vvk}-\ref{tab.HV1345vvk}. We note that one point on 
the $(V-K)$ curve for HV~1335 at phase 0.979 is significant fainter than
the neighbouring points. We assume that this is a deviant point and
not an indication of a possible phase mismatch. In the Baade-Wesslink
analysis we will anyway disregard the phase region from 0.8 to 1.0 as
will be discussed in Storm et al. (\cite{Storm03}).

\section{The Radial Velocities}

  The radial velocity curves were determined from Echelle spectra
obtained with the Las Campanas Observatory 2.5m Du Pont telescope
equipped with the echelle spectrograph and the 2D-Frutti photon counting
system.

The spectrograph provides a dispersion of 0.2nm/mm ($\approx
0.006\mbox{nm~pix}^{-1}$) and was used with a slit of $1.5 \times 4$~arcsec
which transforms into an instrumental profile with a
FWHM of approximately 3.5~pixels or 12~\kms.
The spectrograph is very efficient for our purpose, providing
typical signal to noise ratios of 10 per resolution element
with exposure times ranging from 30 to 60 minutes depending on
the brightness of the object. 

  Additional measurements were secured with the ESO Multi-Mode
Instrument (EMMI) mounted at the ESO New Technology Telescope (NTT) at the
La Silla Observatory. The red arm of the instrument with a CCD detector
(Loral \#34) was used with the echelle grating \#10 giving a dispersion
of 0.24~nm/mm. The efficiency was not quite as high as for the Las
Campanas instrument, but similar S/N ratios could be obtained with
similar exposure times.

\subsection{Observations}

  The Las Campanas data were obtained during four observing runs in the
seasons 1987, 1988, 1989 and 1994 and comprise 171 science spectra
for the objects described here. The La Silla data were obtained during
technical time on one night in October 1992 and a few nights in September
1993 and add an
additional 8 velocity measurements. The exact timing of the exposures
is listed together with the resulting radial velocities in
Table \ref{tab.smcrv} and \ref{tab.lmcrv}.

  Reference Th-Ar exposures of 600~seconds were obtained just before or
after each
of the scientific exposures.  As radial velocity reference we observed
the twilight sky as well as the star C349 in 47~Tuc (chart in Chun
\& Freeman, \cite{Chun78}). 47~Tuc is conveniently located near the SMC
so the instrument is pointing in almost the same direction as during the
science exposures, thus minimizing any differential flexure and ensuring
a minimum overhead of observing time.

\subsection{Data Reduction}

  The echelle spectra were reduced and analysed using IRAF.

  The raw data frames were flat-fielded using an average flat-field for
each observing run which had the large-scale structure removed and which
was normalized. In this way the statistical properties of the data
were largely maintained.

The IRAF package
{\tt noao.imred.echelle} was used for extracting the orders and for
performing the wavelength calibration. Due to limited computing power
at the beginning of the project, 10 orders covering the range
$\lambda\lambda~475-555$~nm
were used for the first three runs and for the Nov. 1994 run 20 orders
covering the range $\lambda\lambda~417-555$~nm were used. For the NTT
data the range $\lambda\lambda~490-640$~nm was used.

  Th-Ar exposures obtained immediately before or immediately after the
science exposure with the telescope in the same position
were employed for defining the wavelength solution.

  The resulting one-dimensional spectra were then fed into the
cross-correlation program {\tt XCOR} (for a description see 
Morse et al. \cite{Morse91})
and for the later runs the related program {\tt XCSAO}
(Kurtz et al. \cite{Kurtz92}) which is part of the {\tt RVSAO} package for IRAF.
The algorithms are based on the technique described by Tonry \& Davis 
(\cite{Tonry79}).

  Each echelle order had the continuum fitted and removed by a 5 piece
cubic spline and the spectra were rebinned to 2048 bins before the
actual cross-correlation.

  Extensive tests were performed to find the best wavelength range and
to choose the optimal 
parameters for the filtering of the Fourier spectrum. 
We found that the best wavelength region was 410-555~nm
so these region was used for the last (1994) run, whereas for the
earlier runs only the region 475-555~nm was employed. The best filter
turned out to ramp up from wavenumber 30 to 40 and taper off from 
350 to 400.

In general the twilight sky exposure gave the best cross-correlation
peak due to a better signal to noise ratio, and 47~Tuc-C349 was used
to determine nightly velocity offsets.  For the early data a reference
velocity of $-21.5 \pm 0.2$~\kms was adopted on the basis of the average
values from the 1989 run itself and from same epoch CORAVEL data (see
Storm et al., \cite{Storm91}).  The mean radial velocity for the 1994
run came out as $-20.8 \pm 0.6$~\kms for
47~Tuc-C349 but this is mostly due to a different selection of wavelength
range. Using the same wavelength range as the previous years gave
$-21.8$~\kms. As we are interested in the best possible relative
velocities between the different observing runs we have adopted the
value of $-21.5$~\kms as the reference velocity and applied nightly
shifts to all the nights to bring all the velocities on the same
system.

  All velocities are referred to the barycenter of the solar system using
the facility within {\tt XCSAO} for transferring the observed velocities to
this reference frame.

\begin{figure}[htp]
\epsfig{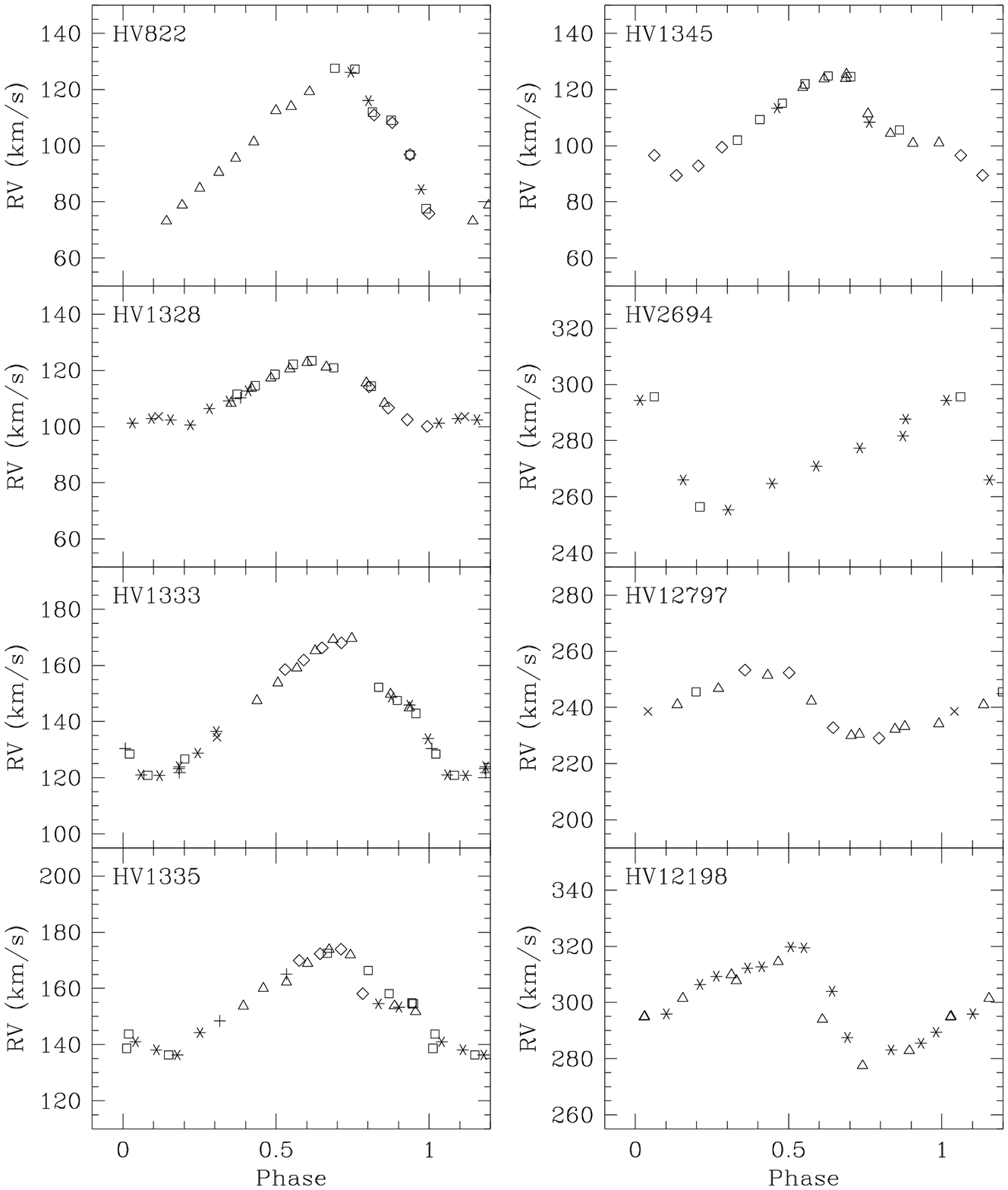}
\centering
\figcap{\label{fig.rv}Radial velocity curves for the cepheids. Note that
the radial velocity scale is the same in all the frames. The data are
labelled according to the run they were aquired. Open boxes are from 1987,
open diamonds 1988, open triangles 1989, asterisks 1994 and crosses and
plusses are NTT data from 1992 and 1993 respectively.
The three last stars are LMC stars.}
\end{figure}

  The final velocities were determined from weighted means of the
velocities determined for the individual orders. The typical standard
deviation for a single order is 3~\kms and using 10 or 20 orders thus
brings down the formal error on the mean to well below 1~\kms. To this
error has to be added the error on the nightly shifts as determined from
the observations of C349. This error can be estimated from the standard
deviations from the mean and is of the order 1.4~\kms which should be
added quadratically to the internal formal error. Realistic
errors can also be estimated from the radial velocity curves themselves
(see Fig.\ref{fig.rv}) by determining the spread around a
smooth curve. We estimate a standard error of 1.5-2.0~\kms per
observation which is also in good agreement with the computed value.

  The velocities are listed in Table \ref{tab.smcrv}
and \ref{tab.lmcrv} together with the Heliocentric Julian Dates (HJD) at
mid-exposure. In Fig.\ref{fig.rv} the radial velocity curves resulting
from these observations and adopting the ephemerides from
Table \ref{tab.ephem} are plotted.

\begin{table}
\caption{\label{tab.C349rv} Radial velocities for the reference star
47~Tuc-C349 measured with respect to the twilight sky.}
\begin{tabular}{r r}
\hline\hline
\multicolumn{1}{c}{HJD} & Rad. vel. \\
 & \kms \\
\hline
2449234.7919 & $-20.6$  \\
2449237.7682 & $-20.3$  \\
2449672.5341 & $-18.3$  \\
2449673.5209 & $-22.2$  \\
2449674.5146 & $-20.8$  \\
2449675.5215 & $-18.5$  \\
2449676.5149 & $-21.6$  \\
2449678.5130 & $-19.9$  \\
2449679.5442 & $-22.0$  \\
\hline
\end{tabular}
\end{table}


A few measurements for \object{HV~821} and \object{HV~883} were
obtained as well and they are listed in Table \ref{tab.addrv}.

\section{Observational Parameters}


  In Table \ref{tab.meanBVIK} we summarize the integrated properties of the
stars. The magnitudes are all intensity magnitude averages. To assure
the best possible phase coverage and optimal definition of the mean
magnitudes we have performed the $V$ and $I$ averages over the data
presented here combined with the Udalski et al. (\cite{Udalski99}) data for
\object{HV~822}, \object{HV~1335}, and \object{HV~1345}, and the
Caldwell \& Coulson (\cite{CC84}) data, offset by $+0.07$~mag in $V$,
for \object{HV~1335}. 
The reddening insensitive Wesenheit index, $\langle W \rangle$, defined
as $\langle W \rangle = \langle V \rangle - 2.51 (\langle V \rangle -
\angle I \rangle$ where
sharp brackets indicates intensity averages, has been tabulated as well.

\begin{table*}
\caption{\label{tab.meanBVIK} Intensity mean magnitudes and
colors based on the combined datasets for each star in the sample.}
\begin{tabular}{r c c c c c c c}
\hline\hline
ID & $\langle V \rangle$ & $\langle B \rangle - \langle V \rangle$ &
$\langle V \rangle - \langle R \rangle$ & 
$\langle V \rangle - \langle I \rangle$ & 
$\langle V \rangle - \langle K \rangle$ &
$\langle J \rangle - \langle K \rangle$
& $\langle W \rangle$ \\
\hline
     HV822 & 14.523 & 0.769 & 0.409 & 0.861 & 1.903 & 0.482 & 12.362\\
    HV1328 & 14.116 & 0.566 & 0.364 & 0.678 & 1.538 & 0.407 & 12.413\\
    HV1333 & 14.707 & 0.712 & 0.432 & 0.839 & 1.905 & 0.472 & 12.601\\
    HV1335 & 14.808 & 0.648 & 0.440 & 0.798 & 1.823 & 0.419 & 12.805\\
    HV1345 & 14.763 & 0.694 & 0.397 & 0.800 & 1.924 & 0.499 & 12.756\\
\hline
\end{tabular}
\end{table*}


\begin{table}
\caption{\label{tab.fourierV} The Fourier coefficients for the $V$
light curves for the SMC Cepheids.}
\scriptsize
\begin{tabular}{l c r r r r r}
\hline\hline
ID & $A_1$ & $k$ & $R_{k1}$ & $\sigma(R_{k1})$ & $\phi_{k1}$ & $\sigma(\phi_{k1})$ \\
   & mag  &     &          &                  &             &   \\
\hline
 & & & & & & \\
     HV822 &   0.41 &  2 & 0.31 & 0.04 & 4.51 & 0.12 \\
           &        &  3 & 0.18 & 0.04 & 2.58 & 0.23 \\
           &        &  4 & 0.11 & 0.03 & 1.15 & 0.36 \\
 & & & & & & \\
    HV1328 &   0.32 &  2 & 0.19 & 0.02 & 4.33 & 0.10 \\
           &        &  3 & 0.14 & 0.02 & 1.58 & 0.14 \\
           &        &  4 & 0.11 & 0.02 & 5.28 & 0.17 \\
 & & & & & & \\
    HV1333 &   0.38 &  2 & 0.37 & 0.04 & 4.88 & 0.15 \\
           &        &  3 & 0.22 & 0.05 & 2.79 & 0.22 \\
           &        &  4 & 0.26 & 0.04 & 1.42 & 0.24 \\
 & & & & & & \\
    HV1335 &   0.34 &  2 & 0.30 & 0.04 & 4.39 & 0.18 \\
           &        &  3 & 0.13 & 0.04 & 1.35 & 0.39 \\
           &        &  4 & 0.08 & 0.04 & 5.33 & 0.51 \\
 & & & & & & \\
    HV1345 &   0.32 &  2 & 0.20 & 0.04 & 4.49 & 0.18 \\
           &        &  3 & 0.18 & 0.04 & 1.87 & 0.22 \\
           &        &  4 & 0.15 & 0.04 & 5.50 & 0.26 \\
\hline
\end{tabular}
\end{table}

\begin{table}
\caption{\label{tab.fourierRV} The Fourier coefficients for the radial
velocity curves for the SMC Cepheids.}
\scriptsize
\begin{tabular}{l c r r r r r}
\hline\hline
ID & $A_1$ & $k$ & $R_{k1}$ & $\sigma(R_{k1})$ & $\phi_{k1}$ & $\sigma(\phi_{k1})$ \\
   & \kms  &     &          &                  &             &   \\
\hline
 & & & & & & \\
     HV822 &     27 &  2 & 0.25 & 0.03 & 2.89 & 0.14 \\
           &        &  3 & 0.09 & 0.03 & 4.62 & 0.32 \\
           &        &  4 & 0.14 & 0.03 & 0.42 & 0.24 \\
 & & & & & & \\
    HV1328 &     11 &  2 & 0.19 & 0.02 & 4.17 & 0.12 \\
           &        &  3 & 0.10 & 0.02 & 0.14 & 0.21 \\
           &        &  4 & 0.08 & 0.02 & 2.31 & 0.27 \\
 & & & & & & \\
    HV1333 &     22 &  2 & 0.13 & 0.02 & 2.93 & 0.18 \\
           &        &  3 & 0.08 & 0.02 & 3.63 & 0.27 \\
           &        &  4 & 0.09 & 0.02 & 0.17 & 0.26 \\
 & & & & & & \\
    HV1335 &     16 &  2 & 0.06 & 0.05 & 3.05 & 0.86 \\
           &        &  3 & 0.13 & 0.05 & 3.23 & 0.42 \\
           &        &  4 & 0.11 & 0.05 & 0.23 & 0.46 \\
 & & & & & & \\
    HV1345 &     15 &  2 & 0.18 & 0.05 & 5.40 & 0.28 \\
           &        &  3 & 0.22 & 0.05 & 2.54 & 0.26 \\
           &        &  4 & 0.07 & 0.05 & 5.94 & 0.70 \\
\hline
\end{tabular}
\end{table}

\begin{figure}[htp]
\epsfig{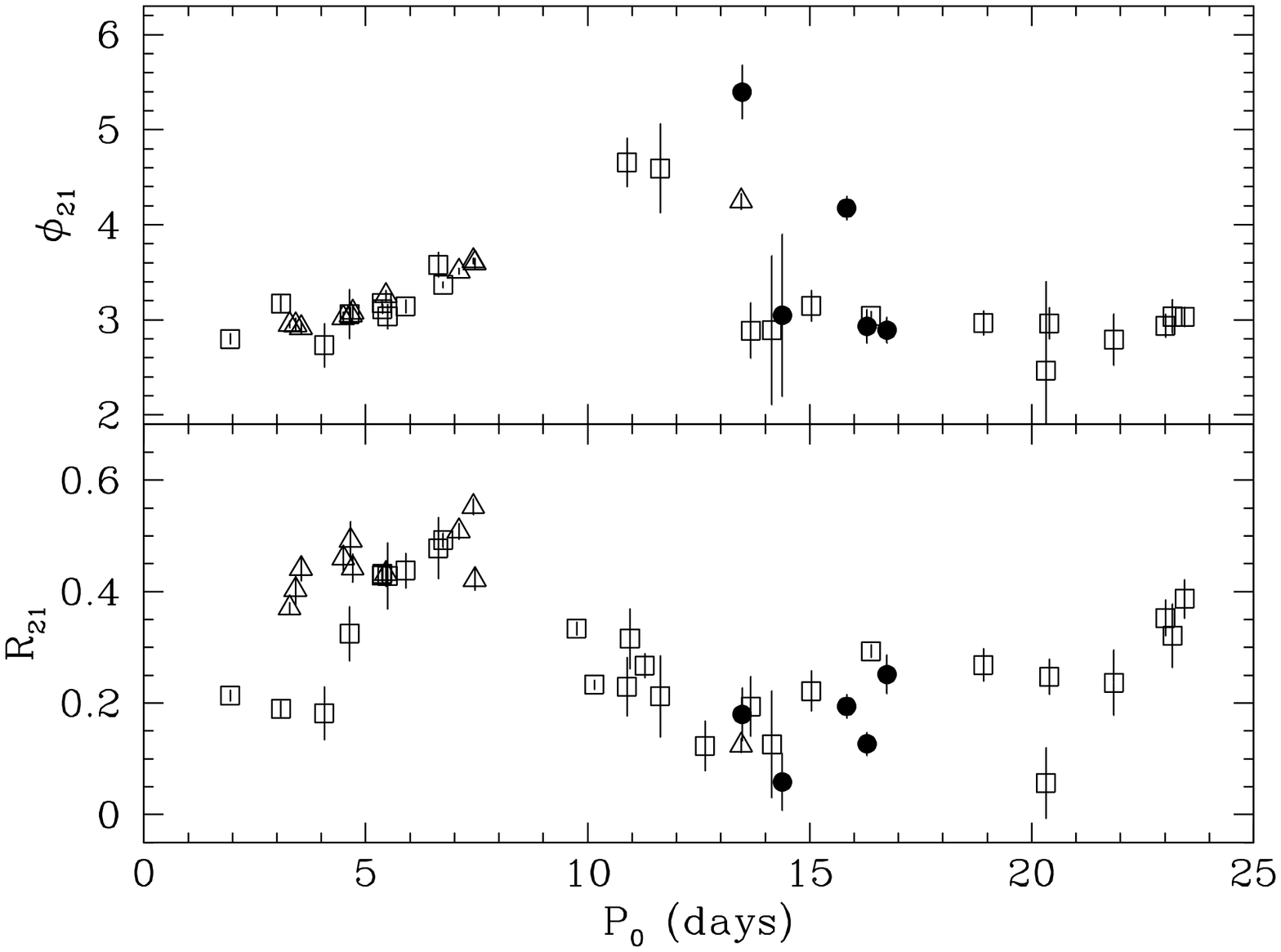}
\figcap{\label{fig.PphiR} The radial velocity Fourier parameters for
our program stars (filled circles), a solar metallicity galactic
sample (open squares), and the low metallicity galactic sample
of Pont et al. (\cite{Pont01}) (open triangles).}
\end{figure}

  We have also computed the Fourier parameters $R_{21}$ - $R_{41}$ and
$\phi_{21}$ - $\phi_{41}$ (as defined by Petersen (\cite{Petersen86}))
for the Cepheids based on the $V$ light curves using $N=4$, i.e. four
components.  The resulting parameters are tabulated in Table \ref{tab.fourierV}
together with the associated error estimates. The semi-amplitudes for
the first ($A_1$) order are also tabulated.  Table \ref{tab.fourierRV}
contains the same parameters but derived for the radial velocity curves
and following the convention from Simon (\cite{Simon88}) where the
radial velocity coefficients are based on a sine series, which means
that the phases are shifted by $\pi / 2$ with respect to the cosine
series used for the photometric data.

The photometric amplitude ratios $R_{21}$ are in good agreement with
the values found by Udalski et al. (\cite{Udalski99}) for their
large sample of SMC stars.

Pont et al. (\cite{Pont01}) have obtained accurate radial velocity curves
for a sample of Galactic cepheids at relatively large Galactocentric
distances with expected lower than solar metallicities.  They find
weak evidence that the $\phi_{21}$ parameter is metallicity sensitive
for periods longer than about 12~days as the star \object{HW~Pup}
exhibits a significantly larger $\phi_{21}$ than that found for solar
metallicity Cepheids.  \object{HW~Pup} with a period of 13.5~days
is indeed metal-poor, with $\FeH=-0.29$ according to Andrievsky et al.
(\cite{Andrievsky02}) and $-0.51$ according to de~Almeida (priv. comm.).

To further investigate this possibility we have overplotted the data
from Pont et al. (\cite{Pont01}) in Fig.\ref{fig.PphiR} with our data
and data from a sample of nearby galactic stars with good radial
velocity curves (see the companion paper Storm et al. \cite{Storm03}).
We see that a couple of
our stars exhibit an even higher value of $\phi_{21}$ than Pont et al.
(\cite{Pont01}) found for \object{HW~Pup}.  However, we also have three
SMC stars which are in good agreement with the supposedly solar metallicity
galactic sample. A clarification of this question will have to await
individual spectroscopic metallicities for these stars and possibly also
a larger sample.

\section*{Acknowledgements}
We appreciate the help we have received from Laura Kellar Fullton and Ricardo
Mu\~noz in collecting the photometric data. 
BWC thanks the U. S. National Science Foundation for
grants AST-8920742 and AST-9800427 to the University of
North Carolina for support of this research.
WPG acknowledges support for this work from the Chilean 
FONDAP Center for Astrophysics 15010003.
Thanks is also due to J{\o}rgen Otzen Petersen who kindly provided his
IDL program to compute the Fourier coefficients and to the 
staff of the three observatories where the data were obtained.

\section*{References}

\begin{list}{}{\itemindent=-\leftmargin \parsep=0cm}
\raggedright
\bibitem[2002]{Andrievsky02}
Andrievsky, S.M., Kovtyukh, V.V., Luck, R. E., L\'epine, J.R.D., Maciel, W.J., \& Beletsky, Yu. V.; \AandA{2002}{392}{491}
\bibitem[1984]{CC84}
Caldwell, J.A.R., and Coulson, I.M.; 1984, SAAO Circ., 8, 1
\bibitem[1978]{Chun78}
Chun, M.S., \& Freeman, K.C.; \AJ{1978}{83}{376}
\bibitem[2001]{Freedman01}
Freedman, W.L., Madore, B.F., Gibson, B.K., et al.; \ApJ{2001}{553}{47}
\bibitem[1997]{Fouque97}
Fouqu\'e, P., \& Gieren, W.P, \AandA{1997}{320}{799}
\bibitem[1977]{Hodge77}
Hodge, P.W., \& Wright, F.W.; 1977, "The Small Magellanic Cloud",
Univ. of Washington Press, Seattle
\bibitem[1992]{Kurtz92}
Kurtz, M.J., Mink, D.J., Wyatt, W.F., et al.; 1992, in: Worral, D.M., Biemesderfer, C., \& Barnes, J. (eds), PASPC, 25, 432
\bibitem[1992]{Landolt92}
Landolt, A.U.; \AJ{1992}{104}{340}
\bibitem[1994]{LS94}
Laney, C.D., \& Stobie, R.S.; \MNRAS{1994}{266}{441}
\bibitem[1991]{Morse91}
Morse, J.A., Mathieu, R.D., \& Levine, S.E.; \AJ{1991}{101}{1495}
\bibitem[1998]{Persson98}
Persson, S.E., Murphy, D.C., Krzeminski, W., Roth, M., \& Rieke, M.J.; \AJ{1998}{116}{2475}
\bibitem[1986]{Petersen86}
Petersen, J.O.; \AandA{1986}{170}{59}
\bibitem[2001]{Pont01}
Pont, F., Kienzle, F., Gieren, W., \& Fouqu\'e, P.; \AandA{2001}{376}{892}
\bibitem[1993]{Schecter93}
Schecter, P.L., Mateo, M., \& Saha, A.; \PASP{1993}{105}{1342}
\bibitem[1988]{Simon88}
Simon, N.R.; in "Pulsation and Mass Loss in Stars", eds. R. Stalio, L.A.
Willson, Kluwer, Dordrecht, 1988, Astrophys. \& Space Sci. Lib., 148, 27
\bibitem[1987]{Stetson87}
Stetson, P.B.; \PASP{1987}{99}{191}
\bibitem[1991]{Storm91}
Storm, J., Carney, B.W., Freedman, W.L., \& Madore, B.F.; \PASP{1991}{103}{261}
\bibitem[1998]{Storm98}
Storm, J., Carney, B.W., \& Fry, A.M.; in "Views on Distance
Indicators", eds. M. Arnaboldi, F. Capuot, \& A. Rifatto, 1998, Mem.
Soc. Ast. It., 69, 79
\bibitem[1999]{Storm99}
Storm, J., Carney, B.W., \& Fry, A.M.; in "Harmonizing Cosmic Distance
Scales in a Post-HIPPARCOS Era", eds. D. Egret, \& A. Heck , 1999, ASP. Conf.
Series, 167, 320
\bibitem[2000]{Storm00}
Storm, J., Carney, B.W., Gieren, W.P., Fouqu\'e, P., \& Fry, A.M.; in
"The Impact of Large-Scale Surveys on Pulsating Star Research", eds. L.
Szabados, \& D. Kurtz, 2000, ASP Conf. Series, 203, 145
\bibitem[2003]{Storm03}
Storm, J., Carney, B.W., Gieren, W.P., Fry, A.M., Latham, D.W., \& Fouqu\'e, P.; \AandA{2003}{Submitted}{}
\bibitem[1979]{Tonry79}
Tonry, J.L., \& Davis, M.; \AJ{1979}{84}{1511}
\bibitem[1999]{Udalski99}
Udalski, A., Soszy\'nski, I., Szyma\'nski, M., et al.; \Acta{1999}{49}{437}
\bibitem[1987]{Welch87}
Welch, D.L., McLaren, R.A., Madore, B.F., \& McAlary, C.W.; \ApJ{1987}{162}{1987}
\end{list}

\clearpage
\newpage

\begin{table*}
\caption{\label{tab.HV822} Photometric values for \object{HV~822}.}
\scriptsize
\begin{tabular}{l l l l l l l l l l}
\hline\hline
\multicolumn{1}{c}{HJD -} &  & & & & & & & & \\
\multicolumn{1}{c}{2400000.} & \multicolumn{1}{c}{Phase} & 
\multicolumn{1}{c}{$V$} & \multicolumn{1}{c}{$\sigma_V$} &
\multicolumn{1}{c}{$(B - V)$} & \multicolumn{1}{c}{$\sigma_{(B-V)}$} &
\multicolumn{1}{c}{$(V - R)$} & \multicolumn{1}{c}{$\sigma_{(V-R)}$} &
\multicolumn{1}{c}{$(V - I)$} & \multicolumn{1}{c}{$\sigma_{(V-I)}$} \\
\multicolumn{1}{c}{days} &      & \multicolumn{1}{c}{mag} &
\multicolumn{1}{c}{mag} & \multicolumn{1}{c}{mag} &
\multicolumn{1}{c}{mag} & \multicolumn{1}{c}{mag} &
\multicolumn{1}{c}{mag} & \multicolumn{1}{c}{mag} &
\multicolumn{1}{c}{mag} \\
\hline\\
47487.62381 & 0.103 & 14.158 & 0.023 & 0.596 & 0.029 & 0.313 & 0.030 & 0.746 & 0.033 \\
47489.59791 & 0.221 & 14.309 & 0.020 & 0.749 & 0.029 & 0.431 & 0.023 & 0.778 & 0.021 \\
47490.62079 & 0.282 & 14.418 & 0.022 & 0.718 & 0.030 & 0.495 & 0.025 & 0.928 & 0.026 \\
47491.62283 & 0.342 & 14.541 & 0.026 & 0.805 & 0.032 & 0.504 & 0.040 & 0.940 & 0.031 \\
\hline
\end{tabular}
\end{table*}

\begin{table*}
\caption{\label{tab.HV1328} Photometric values for \object{HV~1328}.}
\scriptsize
\begin{tabular}{l l l l l l l l l l}
\hline\hline
\multicolumn{1}{c}{HJD -} &  & & & & & & & & \\
\multicolumn{1}{c}{2400000.} & \multicolumn{1}{c}{Phase} & 
\multicolumn{1}{c}{$V$} & \multicolumn{1}{c}{$\sigma_V$} &
\multicolumn{1}{c}{$(B - V)$} & \multicolumn{1}{c}{$\sigma_{(B-V)}$} &
\multicolumn{1}{c}{$(V - R)$} & \multicolumn{1}{c}{$\sigma_{(V-R)}$} &
\multicolumn{1}{c}{$(V - I)$} & \multicolumn{1}{c}{$\sigma_{(V-I)}$} \\
\multicolumn{1}{c}{days} &      & \multicolumn{1}{c}{mag} &
\multicolumn{1}{c}{mag} & \multicolumn{1}{c}{mag} &
\multicolumn{1}{c}{mag} & \multicolumn{1}{c}{mag} &
\multicolumn{1}{c}{mag} & \multicolumn{1}{c}{mag} &
\multicolumn{1}{c}{mag} \\
\hline\\
47487.56801 & 0.055 & 13.837 & 0.027 & 0.382 & 0.039 & 0.324 & 0.030 & --- & --- \\
47488.54714 & 0.117 & 13.969 & 0.034 & 0.444 & 0.041 & 0.355 & 0.034 & 0.648 & 0.041 \\
47489.55266 & 0.180 & 13.971 & 0.032 & 0.543 & 0.029 & 0.348 & 0.019 & 0.653 & 0.024 \\
47490.58392 & 0.245 & 14.072 & 0.030 & 0.577 & 0.024 & 0.395 & 0.017 & 0.703 & 0.024 \\
\hline
\end{tabular}
\end{table*}

\begin{table*}
\caption{\label{tab.HV1333} Photometric values for \object{HV~1333}.}
\scriptsize
\begin{tabular}{l l l l l l l l l l}
\hline\hline
\multicolumn{1}{c}{HJD -} &  & & & & & & & & \\
\multicolumn{1}{c}{2400000.} & \multicolumn{1}{c}{Phase} & 
\multicolumn{1}{c}{$V$} & \multicolumn{1}{c}{$\sigma_V$} &
\multicolumn{1}{c}{$(B - V)$} & \multicolumn{1}{c}{$\sigma_{(B-V)}$} &
\multicolumn{1}{c}{$(V - R)$} & \multicolumn{1}{c}{$\sigma_{(V-R)}$} &
\multicolumn{1}{c}{$(V - I)$} & \multicolumn{1}{c}{$\sigma_{(V-I)}$} \\
\multicolumn{1}{c}{days} &      & \multicolumn{1}{c}{mag} &
\multicolumn{1}{c}{mag} & \multicolumn{1}{c}{mag} &
\multicolumn{1}{c}{mag} & \multicolumn{1}{c}{mag} &
\multicolumn{1}{c}{mag} & \multicolumn{1}{c}{mag} &
\multicolumn{1}{c}{mag} \\
\hline\\
47487.58486 & 0.772 & 15.067 & 0.015 & 0.868 & 0.033 & 0.475 & 0.027 & 0.977 & 0.022 \\
47489.57384 & 0.894 & 15.022 & 0.014 & 0.799 & 0.036 & 0.414 & 0.024 & 0.872 & 0.040 \\
47490.59735 & 0.957 & 14.401 & 0.011 & 0.464 & 0.031 & 0.345 & 0.020 & 0.675 & 0.025 \\
47491.66814 & 0.023 & 14.296 & 0.014 & 0.481 & 0.027 & 0.324 & 0.018 & 0.659 & 0.021 \\
\hline
\end{tabular}
\end{table*}

\begin{table*}
\caption{\label{tab.HV1335} Photometric values for \object{HV~1335}.}
\scriptsize
\begin{tabular}{l l l l l l l l l l}
\hline\hline
\multicolumn{1}{c}{HJD -} &  & & & & & & & & \\
\multicolumn{1}{c}{2400000.} & \multicolumn{1}{c}{Phase} & 
\multicolumn{1}{c}{$V$} & \multicolumn{1}{c}{$\sigma_V$} &
\multicolumn{1}{c}{$(B - V)$} & \multicolumn{1}{c}{$\sigma_{(B-V)}$} &
\multicolumn{1}{c}{$(V - R)$} & \multicolumn{1}{c}{$\sigma_{(V-R)}$} &
\multicolumn{1}{c}{$(V - I)$} & \multicolumn{1}{c}{$\sigma_{(V-I)}$} \\
\multicolumn{1}{c}{days} &      & \multicolumn{1}{c}{mag} &
\multicolumn{1}{c}{mag} & \multicolumn{1}{c}{mag} &
\multicolumn{1}{c}{mag} & \multicolumn{1}{c}{mag} &
\multicolumn{1}{c}{mag} & \multicolumn{1}{c}{mag} &
\multicolumn{1}{c}{mag} \\
\hline\\
47487.58486 & 0.844 & 14.931 & 0.011 & 0.610 & 0.025 & 0.432 & 0.017 & 0.732 & 0.019 \\
47489.57384 & 0.983 & 14.398 & 0.011 & 0.444 & 0.033 & 0.269 & 0.020 & 0.548 & 0.039 \\
47490.59735 & 0.054 & 14.390 & 0.009 & 0.464 & 0.029 & 0.329 & 0.017 & 0.618 & 0.024 \\
47491.66814 & 0.128 & 14.489 & 0.011 & 0.563 & 0.024 & 0.359 & 0.016 & 0.622 & 0.018 \\
\hline
\end{tabular}
\end{table*}

\begin{table*}
\caption{\label{tab.HV1345} Photometric values for \object{HV~1345}.}
\scriptsize
\begin{tabular}{l l l l l l l l l l}
\hline\hline
\multicolumn{1}{c}{HJD -} &  & & & & & & & & \\
\multicolumn{1}{c}{2400000.} & \multicolumn{1}{c}{Phase} & 
\multicolumn{1}{c}{$V$} & \multicolumn{1}{c}{$\sigma_V$} &
\multicolumn{1}{c}{$(B - V)$} & \multicolumn{1}{c}{$\sigma_{(B-V)}$} &
\multicolumn{1}{c}{$(V - R)$} & \multicolumn{1}{c}{$\sigma_{(V-R)}$} &
\multicolumn{1}{c}{$(V - I)$} & \multicolumn{1}{c}{$\sigma_{(V-I)}$} \\
\multicolumn{1}{c}{days} &      & \multicolumn{1}{c}{mag} &
\multicolumn{1}{c}{mag} & \multicolumn{1}{c}{mag} &
\multicolumn{1}{c}{mag} & \multicolumn{1}{c}{mag} &
\multicolumn{1}{c}{mag} & \multicolumn{1}{c}{mag} &
\multicolumn{1}{c}{mag} \\
\hline\\
47487.60924 & 0.377 & 14.787 & 0.022 & 0.847 & 0.034 & 0.405 & 0.032 & 0.846 & 0.039 \\
47488.58073 & 0.450 & 14.978 & 0.036 & 0.834 & 0.053 & 0.472 & 0.038 & 0.956 & 0.085 \\
47489.58555 & 0.524 & 15.017 & 0.021 & 0.877 & 0.028 & 0.462 & 0.021 & 0.910 & 0.030 \\
47490.60981 & 0.600 & 15.109 & 0.018 & 0.872 & 0.023 & 0.485 & 0.019 & 0.948 & 0.026 \\
\hline
\end{tabular}
\end{table*}

\clearpage

\begin{table}
\scriptsize
\caption{\label{tab.HV822IR} IR photometry for \object{HV~822}}
\begin{tabular}{l l l l l l}
\hline\hline
\multicolumn{1}{c}{HJD} & \multicolumn{1}{c}{Phase} & \multicolumn{1}{c}{$K$}
 & \multicolumn{1}{c}{$\sigma (K)$} & \multicolumn{1}{c}{$(J-K)$} &
\multicolumn{1}{c}{$\sigma(J - K)$}\\
\multicolumn{1}{c}{$-2400000$} & & & & & \\
\hline
50820.54459 & 0.177 & 12.529 & 0.029 & 0.369 & 0.033\\
50821.51467 & 0.235 & 12.504 & 0.004 & 0.435 & 0.005\\
50822.52278 & 0.295 & 12.502 & 0.007 & 0.427 & 0.016\\
50823.51922 & 0.355 & 12.455 & 0.003 & 0.449 & 0.004\\
50824.51542 & 0.414 & 12.453 & 0.005 & 0.477 & 0.011\\
51089.61935 & 0.249 & 12.484 & 0.014 & 0.452 & 0.020\\
51091.56660 & 0.365 & 12.465 & 0.023 & 0.482 & 0.033\\
51092.54709 & 0.424 & 12.485 & 0.022 & 0.498 & 0.031\\
51092.65820 & 0.430 & 12.491 & 0.016 & 0.482 & 0.023\\
51093.56861 & 0.485 & 12.489 & 0.023 & 0.528 & 0.033\\
 & & & & & \\
51093.64631 & 0.489 & 12.523 & 0.014 & 0.514 & 0.020\\
51094.54704 & 0.543 & 12.532 & 0.016 & 0.555 & 0.023\\
51094.63107 & 0.548 & 12.568 & 0.008 & 0.541 & 0.011\\
51174.67654 & 0.329 & 12.460 & 0.005 & 0.456 & 0.012\\
51175.59593 & 0.384 & 12.451 & 0.022 & 0.497 & 0.031\\
51176.59805 & 0.444 & 12.461 & 0.005 & 0.526 & 0.006\\
51183.66462 & 0.866 & 12.861 & 0.018 & 0.446 & 0.025\\
51185.61787 & 0.983 & 12.743 & 0.023 & 0.349 & 0.033\\
51205.56010 & 0.174 & 12.534 & 0.006 & 0.434 & 0.007\\
51206.57404 & 0.235 & 12.506 & 0.004 & 0.427 & 0.005\\
 & & & & & \\
51207.56762 & 0.294 & 12.455 & 0.002 & 0.456 & 0.009\\
51208.57264 & 0.354 & 12.453 & 0.024 & 0.480 & 0.034\\
51566.57520 & 0.737 & 12.722 & 0.026 & 0.611 & 0.037\\
51567.51820 & 0.794 & 12.803 & 0.011 & 0.565 & 0.016\\
51568.53480 & 0.854 & 12.841 & 0.020 & 0.519 & 0.028\\
51569.52550 & 0.914 & 12.888 & 0.012 & 0.509 & 0.017\\
\hline
\end{tabular}
\end{table}

\begin{table}
\caption{\label{tab.HV1328IR} IR photometry for \object{HV~1328}}
\scriptsize
\begin{tabular}{l l l l l l}
\hline\hline
\multicolumn{1}{c}{HJD} & \multicolumn{1}{c}{Phase} & \multicolumn{1}{c}{$K$}
 & \multicolumn{1}{c}{$\sigma (K)$} & \multicolumn{1}{c}{$(J-K)$} &
\multicolumn{1}{c}{$\sigma(J - K)$}\\
\multicolumn{1}{c}{$-2400000$} & & & & & \\
\hline
50820.55702 & 0.524 & 12.593 & 0.018 & 0.446 & 0.023\\
50821.53085 & 0.585 & 12.686 & 0.047 & 0.434 & 0.056\\
50822.53509 & 0.649 & 12.670 & 0.029 & 0.469 & 0.041\\
50823.53067 & 0.712 & 12.692 & 0.051 & 0.414 & 0.061\\
51089.55144 & 0.510 & 12.581 & 0.042 & 0.466 & 0.059\\
51091.52972 & 0.635 & 12.654 & 0.033 & 0.457 & 0.047\\
51092.50612 & 0.697 & 12.668 & 0.036 & 0.487 & 0.051\\
51092.62417 & 0.704 & 12.659 & 0.036 & 0.409 & 0.051\\
51093.52836 & 0.761 & 12.682 & 0.040 & 0.431 & 0.057\\
51093.61845 & 0.767 & 12.733 & 0.042 & 0.351 & 0.059\\
 & & & & & \\
51094.51286 & 0.823 & 12.695 & 0.044 & 0.391 & 0.062\\
51094.60046 & 0.829 & 12.651 & 0.030 & 0.378 & 0.042\\
51174.63282 & 0.883 & 12.632 & 0.014 & 0.416 & 0.015\\
51175.55282 & 0.941 & 12.571 & 0.029 & 0.313 & 0.031\\
51176.54591 & 0.004 & 12.568 & 0.029 & 0.255 & 0.032\\
51178.54165 & 0.130 & 12.527 & 0.010 & 0.367 & 0.012\\
51180.58887 & 0.259 & 12.496 & 0.014 & 0.435 & 0.018\\
51183.59299 & 0.449 & 12.541 & 0.041 & 0.479 & 0.058\\
51185.55519 & 0.573 & 12.670 & 0.052 & 0.442 & 0.074\\
51204.57956 & 0.774 & 12.683 & 0.024 & 0.396 & 0.029\\
 & & & & & \\
51205.52678 & 0.834 & 12.661 & 0.007 & 0.398 & 0.018\\
51206.53025 & 0.897 & 12.599 & 0.014 & 0.390 & 0.019\\
51207.53372 & 0.960 & 12.498 & 0.034 & 0.370 & 0.048\\
51208.53233 & 0.024 & 12.510 & 0.020 & 0.287 & 0.026\\
\hline
\end{tabular}
\end{table}

\begin{table}
\caption{\label{tab.HV1333IR} IR photometry for \object{HV~1333}}
\scriptsize
\begin{tabular}{l l l l l l}
\hline\hline
\multicolumn{1}{c}{HJD} & \multicolumn{1}{c}{Phase} & \multicolumn{1}{c}{$K$}
 & \multicolumn{1}{c}{$\sigma (K)$} & \multicolumn{1}{c}{$(J-K)$} &
\multicolumn{1}{c}{$\sigma(J - K)$}\\
\multicolumn{1}{c}{$-2400000$} & & & & & \\
\hline
50821.53906 & 0.391 & 12.670 & 0.002 & 0.525 & 0.005\\
50822.54378 & 0.452 & 12.691 & 0.026 & 0.528 & 0.026\\
50823.53940 & 0.513 & 12.744 & 0.012 & 0.534 & 0.023\\
50824.53650 & 0.575 & 12.836 & 0.017 & 0.532 & 0.022\\
51091.54086 & 0.962 & 12.927 & 0.021 & 0.341 & 0.030\\
51092.52414 & 0.022 & 12.822 & 0.054 & 0.295 & 0.076\\
51092.63108 & 0.029 & 12.845 & 0.019 & 0.270 & 0.027\\
51093.54079 & 0.085 & 12.797 & 0.017 & 9.999 & 9.999\\
51093.62551 & 0.090 & 12.783 & 0.021 & 0.366 & 0.030\\
51094.52061 & 0.145 & 12.725 & 0.013 & 0.391 & 0.018\\
 & & & & & \\
51094.61025 & 0.150 & 12.742 & 0.055 & 0.378 & 0.078\\
51174.64533 & 0.062 & 12.773 & 0.012 & 0.382 & 0.040\\
51175.56192 & 0.119 & 12.757 & 0.011 & 0.400 & 0.013\\
51176.56321 & 0.180 & 12.704 & 0.011 & 0.422 & 0.030\\
51178.55004 & 0.302 & 12.661 & 0.008 & 0.480 & 0.010\\
51180.60417 & 0.428 & 12.676 & 0.019 & 0.597 & 0.045\\
51183.60973 & 0.612 & 12.828 & 0.017 & 0.553 & 0.024\\
51184.59200 & 0.673 & 12.900 & 0.024 & 0.520 & 0.034\\
51204.59486 & 0.900 & 13.002 & 0.021 & 0.502 & 0.030\\
51206.53930 & 0.020 & 12.778 & 0.012 & 0.359 & 0.018\\
 & & & & & \\
51207.54277 & 0.081 & 12.788 & 0.015 & 0.408 & 0.048\\
\hline
\end{tabular}
\end{table}

\begin{table}
\caption{\label{tab.HV1335IR} IR photometry for \object{HV~1335}}
\scriptsize
\begin{tabular}{l l l l l l}
\hline\hline
\multicolumn{1}{c}{HJD} & \multicolumn{1}{c}{Phase} & \multicolumn{1}{c}{$K$}
 & \multicolumn{1}{c}{$\sigma (K)$} & \multicolumn{1}{c}{$(J-K)$} &
\multicolumn{1}{c}{$\sigma(J - K)$}\\
\multicolumn{1}{c}{$-2400000$} & & & & & \\
\hline
50821.54829 & 0.666 & 13.032 & 0.024 & 0.483 & 0.027\\
50822.55359 & 0.736 & 13.205 & 0.087 & 0.321 & 0.123\\
50823.54960 & 0.805 & 13.222 & 0.020 & 0.350 & 0.028\\
50824.54488 & 0.874 & 13.051 & 0.032 & 0.504 & 0.043\\
51089.57911 & 0.303 & 12.895 & 0.056 & 0.346 & 0.079\\
51091.54857 & 0.440 & 12.874 & 0.034 & 0.467 & 0.048\\
51092.53111 & 0.508 & 12.927 & 0.015 & 0.458 & 0.021\\
51092.63738 & 0.516 & 12.884 & 0.055 & 0.516 & 0.078\\
51093.55166 & 0.579 & 13.017 & 0.076 & 0.455 & 0.107\\
51093.63177 & 0.585 & 13.084 & 0.051 & 0.369 & 0.072\\
 & & & & & \\
51094.53171 & 0.647 & 13.093 & 0.035 & 0.417 & 0.049\\
51094.61775 & 0.653 & 13.089 & 0.085 & 0.429 & 0.120\\
51174.65707 & 0.219 & 12.856 & 0.060 & 0.425 & 0.085\\
51175.57443 & 0.282 & 12.844 & 0.030 & 0.402 & 0.030\\
51176.57860 & 0.352 & 12.837 & 0.065 & 0.405 & 0.092\\
51178.56675 & 0.491 & 12.907 & 0.040 & 0.449 & 0.052\\
51183.62818 & 0.842 & 13.134 & 0.033 & 0.430 & 0.047\\
51184.60451 & 0.910 & 13.093 & 0.014 & 0.410 & 0.020\\
51204.60181 & 0.301 & 12.872 & 0.059 & 0.384 & 0.059\\
51205.54625 & 0.367 & 12.822 & 0.075 & 0.437 & 0.106\\
 & & & & & \\
51206.55111 & 0.436 & 12.867 & 0.022 & 0.451 & 0.028\\
51207.55181 & 0.506 & 12.898 & 0.019 & 0.483 & 0.033\\
51208.54972 & 0.575 & 13.007 & 0.049 & 0.448 & 0.051\\
\hline
\end{tabular}
\end{table}

\begin{table}
\caption{\label{tab.HV1345IR} IR photometry for \object{HV~1345}}
\scriptsize
\begin{tabular}{l l l l l l}
\hline\hline
\multicolumn{1}{c}{HJD} & \multicolumn{1}{c}{Phase} & \multicolumn{1}{c}{$K$}
 & \multicolumn{1}{c}{$\sigma (K)$} & \multicolumn{1}{c}{$(J-K)$} &
\multicolumn{1}{c}{$\sigma(J - K)$}\\
\multicolumn{1}{c}{$-2400000$} & & & & & \\
\hline
50820.58937 & 0.661 & 12.891 & 0.025 & 0.557 & 0.054\\
50821.55653 & 0.732 & 12.927 & 0.014 & 0.579 & 0.015\\
50822.56058 & 0.807 & 12.984 & 0.006 & 0.506 & 0.011\\
50823.55734 & 0.881 & 13.015 & 0.009 & 0.482 & 0.012\\
50824.55426 & 0.955 & 12.895 & 0.011 & 0.423 & 0.013\\
51089.60282 & 0.619 & 12.874 & 0.016 & 0.546 & 0.023\\
51091.55691 & 0.764 & 13.018 & 0.022 & 0.482 & 0.031\\
51092.53885 & 0.837 & 12.981 & 0.027 & 0.490 & 0.038\\
51092.64998 & 0.845 & 13.031 & 0.023 & 0.448 & 0.033\\
51093.56071 & 0.913 & 12.949 & 0.022 & 0.464 & 0.031\\
 & & & & & \\
51093.63928 & 0.919 & 12.956 & 0.026 & 0.439 & 0.037\\
51094.53946 & 0.986 & 12.832 & 0.020 & 0.372 & 0.028\\
51094.62488 & 0.992 & 12.846 & 0.018 & 0.336 & 0.025\\
51174.66597 & 0.930 & 12.927 & 0.010 & 0.479 & 0.017\\
51175.58481 & 0.999 & 12.866 & 0.011 & 0.403 & 0.013\\
51176.58696 & 0.073 & 12.826 & 0.005 & 0.419 & 0.006\\
51178.57490 & 0.220 & 12.737 & 0.005 & 0.513 & 0.021\\
51183.64428 & 0.597 & 12.880 & 0.015 & 0.537 & 0.021\\
51184.62009 & 0.669 & 12.913 & 0.020 & 0.548 & 0.028\\
51185.60139 & 0.742 & 12.960 & 0.023 & 0.494 & 0.033\\
 & & & & & \\
51205.55181 & 0.222 & 12.726 & 0.014 & 0.503 & 0.015\\
51206.56430 & 0.297 & 12.698 & 0.012 & 0.532 & 0.013\\
51207.56083 & 0.371 & 12.721 & 0.003 & 0.542 & 0.004\\
51208.56083 & 0.445 & 12.757 & 0.003 & 0.539 & 0.010\\
51568.57880 & 0.156 & 12.762 & 0.011 & 0.486 & 0.016\\
51569.54940 & 0.228 & 12.776 & 0.011 & 0.472 & 0.016\\
\hline
\end{tabular}
\end{table}

\clearpage

\begin{table*}
\caption{\label{tab.HV822vvk} The $V$ and $(V-K)$ color curves for
HV~822. The $(V-K)$ color has been determined
on the basis of the observed $V$ data and the fourier fit to the $K$-band data.}
\scriptsize
\begin{tabular}{c c c c c | c c c c c | c c c c c}
\hline\hline
Phase & $V$ & $\sigma (V)$ & $(V-K)$ & $\sigma(V - K)$ &
Phase & $V$ & $\sigma (V)$ & $(V-K)$ & $\sigma(V - K)$ &
Phase & $V$ & $\sigma (V)$ & $(V-K)$ & $\sigma(V - K)$\\
 & mag & mag & mag & mag & 
 & mag & mag & mag & mag & 
 & mag & mag & mag & mag \\
\hline
0.002 & 13.892 & 0.007 &  1.128 & 0.012 &0.274 & 14.313 & 0.007 &  1.836 & 0.012 &0.504 & 14.693 & 0.006 &  2.177 & 0.012 \\
0.036 & 14.010 & 0.008 &  1.311 & 0.013 &0.282 & 14.418 & 0.022 &  1.946 & 0.024 &0.522 & 14.775 & 0.023 &  2.243 & 0.025 \\
0.054 & 14.077 & 0.007 &  1.393 & 0.012 &0.297 & 14.330 & 0.010 &  1.868 & 0.014 &0.539 & 14.721 & 0.010 &  2.173 & 0.014 \\
0.075 & 14.097 & 0.011 &  1.444 & 0.015 &0.300 & 14.346 & 0.010 &  1.884 & 0.014 &0.541 & 14.741 & 0.015 &  2.193 & 0.018 \\
0.103 & 14.158 & 0.023 &  1.545 & 0.025 &0.305 & 14.388 & 0.017 &  1.926 & 0.020 &0.556 & 14.782 & 0.013 &  2.218 & 0.016 \\
0.113 & 14.171 & 0.012 &  1.570 & 0.016 &0.309 & 14.368 & 0.007 &  1.910 & 0.012 &0.582 & 14.807 & 0.022 &  2.228 & 0.024 \\
\hline
\end{tabular}
\end{table*}

\begin{table*}
\caption{\label{tab.HV1328vvk} The $V$ and $(V-K)$ color curves for
HV~1328. The $(V-K)$ color has been determined
on the basis of the observed $V$ data and the fourier fit to the $K$-band data.}
\scriptsize
\begin{tabular}{c c c c c | c c c c c | c c c c c}
\hline\hline
Phase & $V$ & $\sigma (V)$ & $(V-K)$ & $\sigma(V - K)$ &
Phase & $V$ & $\sigma (V)$ & $(V-K)$ & $\sigma(V - K)$ &
Phase & $V$ & $\sigma (V)$ & $(V-K)$ & $\sigma(V - K)$\\
 & mag & mag & mag & mag &
 & mag & mag & mag & mag &
 & mag & mag & mag & mag \\
\hline
0.004 & 13.725 & 0.010 &  1.198 & 0.014 &0.280 & 14.009 & 0.062 &  1.518 & 0.063 &0.638 & 14.473 & 0.009 &  1.811 & 0.013 \\
0.007 & 13.684 & 0.011 &  1.161 & 0.015 &0.303 & 14.158 & 0.019 &  1.673 & 0.021 &0.641 & 14.488 & 0.013 &  1.826 & 0.016 \\
0.013 & 13.750 & 0.008 &  1.227 & 0.013 &0.305 & 14.101 & 0.035 &  1.618 & 0.036 &0.696 & 14.393 & 0.008 &  1.711 & 0.013 \\
0.018 & 13.751 & 0.010 &  1.230 & 0.014 &0.329 & 14.135 & 0.036 &  1.653 & 0.037 &0.701 & 14.388 & 0.008 &  1.706 & 0.013 \\
0.046 & 13.780 & 0.015 &  1.262 & 0.018 &0.349 & 14.135 & 0.037 &  1.652 & 0.038 &0.704 & 14.344 & 0.064 &  1.662 & 0.065 \\
0.049 & 13.781 & 0.013 &  1.263 & 0.016 &0.360 & 14.138 & 0.045 &  1.652 & 0.046 &0.760 & 14.305 & 0.008 &  1.616 & 0.013 \\
\hline
\end{tabular}
\end{table*}

\begin{table*}
\caption{\label{tab.HV1333vvk} The $V$ and $(V-K)$ color curves for
HV~1333. The $(V-K)$ color has been determined
on the basis of the observed $V$ data and the fourier fit to the $K$-band data.}
\scriptsize
\begin{tabular}{c c c c c | c c c c c | c c c c c}
\hline\hline
Phase & $V$ & $\sigma (V)$ & $(V-K)$ & $\sigma(V - K)$ &
Phase & $V$ & $\sigma (V)$ & $(V-K)$ & $\sigma(V - K)$ &
Phase & $V$ & $\sigma (V)$ & $(V-K)$ & $\sigma(V - K)$\\
 & mag & mag & mag & mag & 
 & mag & mag & mag & mag & 
 & mag & mag & mag & mag \\
\hline
0.009 & 14.160 & 0.013 &  1.330 & 0.016 &0.237 & 14.527 & 0.012 &  1.842 & 0.016 &0.572 & 15.049 & 0.030 &  2.252 & 0.032 \\
0.023 & 14.296 & 0.014 &  1.478 & 0.017 &0.258 & 14.502 & 0.015 &  1.827 & 0.018 &0.620 & 15.064 & 0.013 &  2.217 & 0.016 \\
0.048 & 14.303 & 0.006 &  1.515 & 0.012 &0.267 & 14.565 & 0.007 &  1.895 & 0.012 &0.621 & 15.070 & 0.008 &  2.223 & 0.013 \\
0.052 & 14.341 & 0.006 &  1.553 & 0.012 &0.293 & 14.547 & 0.010 &  1.884 & 0.014 &0.632 & 15.048 & 0.033 &  2.190 & 0.034 \\
0.055 & 14.344 & 0.007 &  1.556 & 0.012 &0.294 & 14.574 & 0.010 &  1.911 & 0.014 &0.678 & 15.084 & 0.016 &  2.159 & 0.019 \\
0.055 & 14.322 & 0.009 &  1.534 & 0.013 &0.298 & 14.587 & 0.007 &  1.927 & 0.012 &0.681 & 15.107 & 0.011 &  2.182 & 0.015 \\
\hline
\end{tabular}
\end{table*}

\begin{table*}
\caption{\label{tab.HV1335vvk} The $V$ and $(V-K)$ color curves for
HV~1335. The $(V-K)$ color has been determined
on the basis of the observed $V$ data and the fourier fit to the $K$-band data.}
\scriptsize
\begin{tabular}{c c c c c | c c c c c | c c c c c}
\hline\hline
Phase & $V$ & $\sigma (V)$ & $(V-K)$ & $\sigma(V - K)$ &
Phase & $V$ & $\sigma (V)$ & $(V-K)$ & $\sigma(V - K)$ &
Phase & $V$ & $\sigma (V)$ & $(V-K)$ & $\sigma(V - K)$\\
 & mag & mag & mag & mag &
 & mag & mag & mag & mag &
 & mag & mag & mag & mag \\
\hline
0.002 & 14.445 & 0.041 &  1.492 & 0.042 &0.312 & 14.727 & 0.008 &  1.883 & 0.013 &0.596 & 15.097 & 0.009 &  2.093 & 0.013 \\
0.002 & 14.319 & 0.015 &  1.366 & 0.018 &0.317 & 14.718 & 0.006 &  1.876 & 0.012 &0.597 & 15.255 & 0.084 &  2.251 & 0.085 \\
0.004 & 14.405 & 0.010 &  1.452 & 0.014 &0.320 & 14.730 & 0.005 &  1.888 & 0.011 &0.598 & 15.155 & 0.010 &  2.151 & 0.014 \\
0.009 & 14.305 & 0.010 &  1.362 & 0.014 &0.322 & 14.718 & 0.008 &  1.876 & 0.013 &0.630 & 15.170 & 0.010 &  2.132 & 0.014 \\
0.017 & 14.322 & 0.013 &  1.388 & 0.016 &0.333 & 14.752 & 0.017 &  1.912 & 0.020 &0.656 & 15.135 & 0.010 &  2.060 & 0.014 \\
0.017 & 14.356 & 0.010 &  1.422 & 0.014 &0.333 & 14.731 & 0.013 &  1.891 & 0.016 &0.661 & 15.181 & 0.014 &  2.106 & 0.017 \\
\hline
\end{tabular}
\end{table*}

\begin{table*}
\caption{\label{tab.HV1345vvk} The $V$ and $(V-K)$ color curves for
HV~1345. The $(V-K)$ color has been determined
on the basis of the observed $V$ data and the fourier fit to the $K$-band data.}
\scriptsize
\begin{tabular}{c c c c c | c c c c c | c c c c c}
\hline\hline
Phase & $V$ & $\sigma (V)$ & $(V-K)$ & $\sigma(V - K)$ &
Phase & $V$ & $\sigma (V)$ & $(V-K)$ & $\sigma(V - K)$ &
Phase & $V$ & $\sigma (V)$ & $(V-K)$ & $\sigma(V - K)$\\
 & mag & mag & mag & mag &
 & mag & mag & mag & mag &
 & mag & mag & mag & mag \\
\hline
0.022 & 14.359 & 0.017 &  1.517 & 0.020 &0.408 & 14.900 & 0.015 &  2.164 & 0.018 &0.713 & 15.042 & 0.021 &  2.129 & 0.023 \\
0.025 & 14.348 & 0.011 &  1.506 & 0.015 &0.413 & 14.839 & 0.027 &  2.103 & 0.029 &0.735 & 14.997 & 0.014 &  2.061 & 0.017 \\
0.040 & 14.395 & 0.013 &  1.563 & 0.016 &0.441 & 14.916 & 0.008 &  2.162 & 0.013 &0.739 & 15.003 & 0.010 &  2.067 & 0.014 \\
0.097 & 14.467 & 0.027 &  1.653 & 0.029 &0.449 & 14.928 & 0.011 &  2.167 & 0.015 &0.743 & 15.005 & 0.010 &  2.069 & 0.014 \\
0.099 & 14.471 & 0.014 &  1.657 & 0.017 &0.450 & 14.978 & 0.036 &  2.217 & 0.037 &0.749 & 14.985 & 0.012 &  2.040 & 0.016 \\
0.103 & 14.490 & 0.013 &  1.676 & 0.016 &0.471 & 14.971 & 0.006 &  2.193 & 0.012 &0.752 & 15.095 & 0.054 &  2.150 & 0.055 \\
\hline
\end{tabular}
\end{table*}

\clearpage

\begin{table*}
\caption{\label{tab.smcrv}Radial velocities for the SMC Cepheids. Typical errors, including systematics, are 1.5\kms.}
\footnotesize
\begin{tabular}{r r | r r | r r | r r | r r}
\hline\hline
\multicolumn{2}{c|}{HV~822} & \multicolumn{2}{c|}{HV~1328} &
\multicolumn{2}{c|}{HV~1333} & \multicolumn{2}{c|}{HV~1335} &
\multicolumn{2}{c}{HV~1345}\\
HJD  & Radial &HJD  & Radial &HJD  & Radial &HJD  & Radial &HJD  & Radial \\
$- 2400000 $ & velocity &$- 2400000 $ & velocity &$- 2400000 $ & velocity &$- 2400000 $ & velocity &$- 2400000 $ &
velocity \\
 & \kms & & \kms & & \kms & & \kms & & \kms \\
\hline
47096.5740& 127.6&47096.7153& 111.5&47097.5649& 152.2&47096.6795& 172.5&47096.5365& 102.0\\
47097.6649& 127.2&47097.6319& 114.5&47098.5545& 147.5&47098.5906& 166.4&47097.5295& 109.4\\
47098.6260& 112.0&47098.6558& 118.6&47099.5442& 142.9&47099.5757& 158.2&47098.5191& 115.1\\
47099.6518& 109.1&47099.6007& 122.2&47100.6159& 128.4&47100.6452& 154.8&47099.5148& 122.0\\
\hline
\end{tabular}
\normalsize
\end{table*}

\begin{table*}
\caption{\label{tab.lmcrv}Radial velocities for the LMC Cepheids.}
\footnotesize
\begin{tabular}{r r | r r | r r }
\hline\hline
\multicolumn{2}{c|}{HV~2694} & \multicolumn{2}{c|}{HV~12198} &
\multicolumn{2}{c}{HV~12797}\\
HJD  & Radial &HJD  & Radial &HJD  & Radial \\
$- 2400000 $ & velocity &$- 2400000 $ & velocity &$- 2400000 $ &
velocity \\
\hline
 & \kms & & \kms & & \kms \\
\hline
47100.7531 &  295.6 & 47840.8042 &  295.0 & 47100.8337 &  245.6\\
47101.7882 &  256.4 & 47841.8538 &  307.7 & 47483.8379 &  253.3\\
49672.8161 &  281.7 & 47841.8538 &  307.8 & 47484.8240 &  252.4\\
49673.7967 &  294.4 & 47842.8399 &  294.0 & 47485.8042 &  232.8\\
49674.7752 &  266.0 & 47843.8344 &  282.8 & 47486.8247 &  229.1\\
49675.7953 &  255.3 & 47844.7496 &  301.5 & 47840.8470 &  230.0\\
49676.7857 &  264.8 & 47845.8434 &  314.5 & 47841.8206 &  232.3\\
49677.7898 &  271.0 & 47846.8076 &  277.5 & 47842.7949 &  234.2\\
49678.7724 &  277.4 & 47847.8149 &  294.9 & 47843.7945 &  241.0\\
49679.8078 &  287.7 & 47848.8135 &  309.9 & 47844.7098 &  246.7\\
           &        &            &        &            &       \\
           &        & 49672.7510 &  295.9 & 47845.8053 &  251.5\\
           &        & 49673.6795 &  312.2 & 47846.7799 &  242.3\\
           &        & 49673.8420 &  312.7 & 47847.8507 &  230.5\\
           &        & 49674.6469 &  304.0 & 47848.8577 &  233.3\\
           &        & 49674.8225 &  287.5 & 48913.8813 &  238.6\\
           &        & 49675.6663 &  285.5 &            &         \\
           &        & 49675.8413 &  289.4 &            &         \\
           &        & 49676.6448 &  306.4 &            &         \\
           &        & 49676.8329 &  309.3 &            &         \\
           &        & 49677.6816 &  319.8 &            &         \\
           &        &            &        &            &       \\
           &        & 49677.8364 &  319.5 &            &         \\
           &        & 49678.8336 &  283.1 &            &         \\
\hline
\end{tabular}
\normalsize
\end{table*}

\clearpage

\begin{table}
\caption{\label{tab.addrv} Radial velocities for two additional stars}
\begin{tabular}{c l l}
\hline\hline
ID & HJD & Radial \\
   &$-2400000$ & velocity \\
   &         & \kms \\
\hline
\object{HV~821} (SMC) & 47096.6090 & 100.9 \\
\object{HV~821} (SMC) & 47103.6972 & 105.6 \\
\object{HV~883} (LMC) & 47103.7743 & 243.2 \\
\hline
\end{tabular}
\end{table}

\end{document}